\documentclass[12pt,preprint]{aastex}

\shorttitle{The Globular Cluster System of NGC 4636. II.}
\shortauthors{Lee et al.}

\begin{document}
\title{The Globular Cluster System of the Virgo Giant Elliptical Galaxy NGC 4636:
II. Kinematics of the Globular Cluster System\altaffilmark{*}}  

\author{ Myung Gyoon LEE\altaffilmark{1}, Hong Soo PARK\altaffilmark{1}, 
Ho Seong HWANG\altaffilmark{1,2}, Nobuo Arimoto\altaffilmark{3}, Naoyuki Tamura\altaffilmark{4},
and Masato Onodera\altaffilmark{2} }

\email{mglee@astro.snu.ac.kr, hspark@astro.snu.ac.kr,  hoseong.hwang@cea.fr,
arimoto.n@nao.ac.jp, naoyuki@subaru.naoj.org, masato.onodera@cea.fr}

\altaffiltext{1}{Astronomy Program, Department of Physics and Astronomy, Seoul National University, Korea}
\altaffiltext{2}{CEA, Laboratoire AIM, Irfu/SAp, F-91191 Gif-sur-Yvette, France}
\altaffiltext{3}{National Astronomical Observatory of Japan, Tokyo, Japan}
\altaffiltext{4}{Subaru Telescope, National Astronomical Observatory of Japan, Hilo, USA}
\altaffiltext{*}{Based on data collected at Subaru Telescope, which is operated by the National Astronomical Observatory of Japan.}
 
\begin{abstract}
We present a kinematic analysis of the globular cluster (GC) system in
the giant elliptical galaxy (gE) NGC 4636 in the Virgo cluster.
Using the photometric and spectroscopic database of 238 GCs (108 blue GCs and 130 red GCs) 
  at the galactocentric radius $0\arcmin.39 <R< 15\arcmin.43$, 
  we have investigated the kinematics of the GC system.
The NGC 4636 GC system shows weak overall rotation, which is dominated by the red GCs. 
However, both the blue GCs and red GCs show some rotation in the inner region 
at $R<4\arcmin.3$ ($=2.9 R_{\rm eff} = 18.5$ kpc).
The velocity dispersion for all the GCs is derived to be $\sigma_p = 225^{+12}_{-9}$~ km s$^{-1}$.
The velocity dispersion for the blue GCs ($\sigma_p = 251^{+18}_{-12}$ km s$^{-1}$)
  is slightly larger than that for the red GCs ($\sigma_p = 205^{+11}_{-13}$~ km s$^{-1}$).
The velocity dispersions for the blue GCs about the mean velocity and about 
  the best fit rotation curve have a significant variation depending on the galactocentric radius. 
Comparison of observed stellar and GC velocity dispersion profiles with the velocity dispersion
  profiles calculated from the stellar mass profile
  shows that the mass-to-light ratio should increase as the galactocentric distance increases,
  indicating the existence of an extended dark matter halo.
From the comparison of the observed GC velocity dispersion profiles and the velocity dispersion 
profiles calculated for the X-ray mass profiles in the literature, 
we find that
  the orbit of the GC system is tangential, and that the orbit of 
  the red GCs is slightly more tangential than that of the blue GCs.  
We compare the GC kinematics of NGC 4636  with those of other six gEs,
finding that the kinematic properties of the GCs are diverse among gEs.
We find several correlations between the kinematics of the GCs and the global parameters of their host galaxies. 
We discuss the implication of the results for the formation models of the GC system in gEs,
and suggest a mixture scenario for the origin of the GCs in gEs.

\end{abstract}


\section{Introduction}

Globular clusters (GCs) are an excellent tracer to probe the gravitational potential of their host galaxies.
Kinematics of the GC system in a galaxy is determined by the gravitational potential of their host galaxies
  and the GCs contain a fossil record of the dynamical evolution since the formation of their host galaxy.
From the kinematic study of the GC system
  we can estimate the global mass distribution of their host galaxy including the dark matter,
  or can derive information on the orbital properties of the GCs 
  if a mass distribution of their host galaxy is known in prior (e.g., X-ray observation of the hot gas).
GCs are useful especially for the study of the outer region of a galaxy where the gravitational
  potential is mainly dominated by the dark matter.
They are particularly efficient for the study of nearby giant elliptical galaxies (gEs) 
  where thousands of GCs are often found.

Recently \citet{hwa08} presented a good summary of the results on the kinematics of the GC systems 
  in six nearby gEs, derived using the consistent analysis with the data in the literature:
M49 \citep{zep00,cot03},
M60 \citep{pie06, bri06,lee08a}, 
M87 \citep{coh97,kis98b,cot01}, 
NGC 4636 \citep{sch06, cha08}, 
NGC 1399 \citep{kis98, min98, kis99, ric04, ric08}, and 
NGC 5128 \citep{pen04a, pen04b, woo07}.
Later \citet{rom09} presented the results of a study of the kinematics of the GC system in 
another gE, NGC 1407, 
  in a nearby galaxy group and provided updated kinematic analysis of the GC systems in gEs
  using their consistent method.
One surprising finding from these studies is that the kinematics of the GC systems in these gEs is diverse, 
  showing a large difference in velocity dispersion, rotation, and their radial variation among the gEs. 
This is in stark contrast to the photometric property of the GCs that
  is more or less similar among the gEs \citep{lee03,bro06}.

We have been carrying a project to investigate the spectroscopic properties of the GCs in nearby galaxies. 
Our study on the kinematics of the GC system of M60, a gE in Virgo, was presented in \citet{lee08a} and \citet{hwa08},
and another on the M31 GC system was given in \citet{lee08c}.
Recently we presented the measurement of radial velocities for the GCs in NGC 4636 
  in the companion paper \citep[hereafter Paper I]{par09a}, 
  and we present a detailed kinematic analysis of these data in this paper. 

NGC 4636 is an E/S0 galaxy in Virgo.
It is located $10^{\circ}.8$ (2.8 Mpc at the distance of NGC 4636) south east from the Virgo center, 
  and is considered to be a major member of a small group falling into the Virgo center \citep{nol93,tri94}.
NGC 4636 is relatively less luminous ($M_V=-21.7$  mag) among the gEs in Virgo.
It shows several interesting structures in the central region:
  jets in the radio images, arm-like structures, bubbles, and a cavity in the X-ray images \citep{osu05, bal09},
  and dust emission in the far-infrared (100$\mu m$) image \citep{tem03}.
It is notable that the faint isophotes in the outer region of NGC 4636 are much flatter (E4) than those in the 
  inner region (E0), indicating that this galaxy may be in the transition to S0 \citep{san61}.
This fact may imply the presence of large-scale angular momentum associated with recent mergers.
We adopted a distance to NGC 4636, 
  14.7 Mpc [$(m-M)_0=30.83\pm 0.13$], as given by \citet{ton01}
  based on the surface brightness fluctuation method.
One arcmin in the sky at this distance corresponds to $4.26$ kpc.
Effective radius, ellipticity, and position angle of NGC 4636 are 
$R_{\rm eff}=1.49\arcmin=6.347$ kpc, $\epsilon_{\rm eff} = 0.256$, and PA$_{\rm eff}=148$ deg,
respectively \citep{kim06,par09b}.

This paper is composed as follows.
Section \ref{data} gives a brief description of the data used in this analysis, 
  and the kinematic properties of the GC system in NGC 4636 are derived in \S \ref{results}.
In \S \ref{discuss}, we compare  kinematic properties of the NGC 4636 GC system to those of other gEs,
  and discuss the implication of the results regarding the GC formation models.
Primary results are summarized in the final section.

\section{Data}\label{data}

We used the spectroscopic data of the GCs given in Paper I, 
  which describes the details of the spectroscopic observation, data reduction, and the data set. 
Here we only give a brief summary of the data set of NGC 4636 GCs.
We selected GC candidates in deep, wide-field Washington
  C and $T_1$ images ($15.8\arcmin\times15.8\arcmin$) obtained at KPNO 4m telescope \citep{par09b}.
Spectroscopic observations of these targets were made 
 using the Multi-Object Spectroscopy (MOS) mode of Faint Object Camera and
Spectrograph (FOCAS) on the SUBARU 8.2m Telescope.

We determined the radial velocities for 122 objects including
105 GCs, 11 foreground stars, 2 background galaxies, 3 probable 
intracluster GCs in Virgo, and the nucleus of NGC 4636,
by cross-correlating the target spectra with five Galactic GCs as templates.
Paper I presented a master catalog of the radial velocities for the GCs in NGC 4636 
   combining their data with the data for 174 GCs given by \citet{sch06}. 
The radial velocities of GCs in \citet{sch06} were transformed
into our velocity system using equation (1) in Paper I,
and the transformed velocities were used for further analysis.

\begin{figure}
\plotone{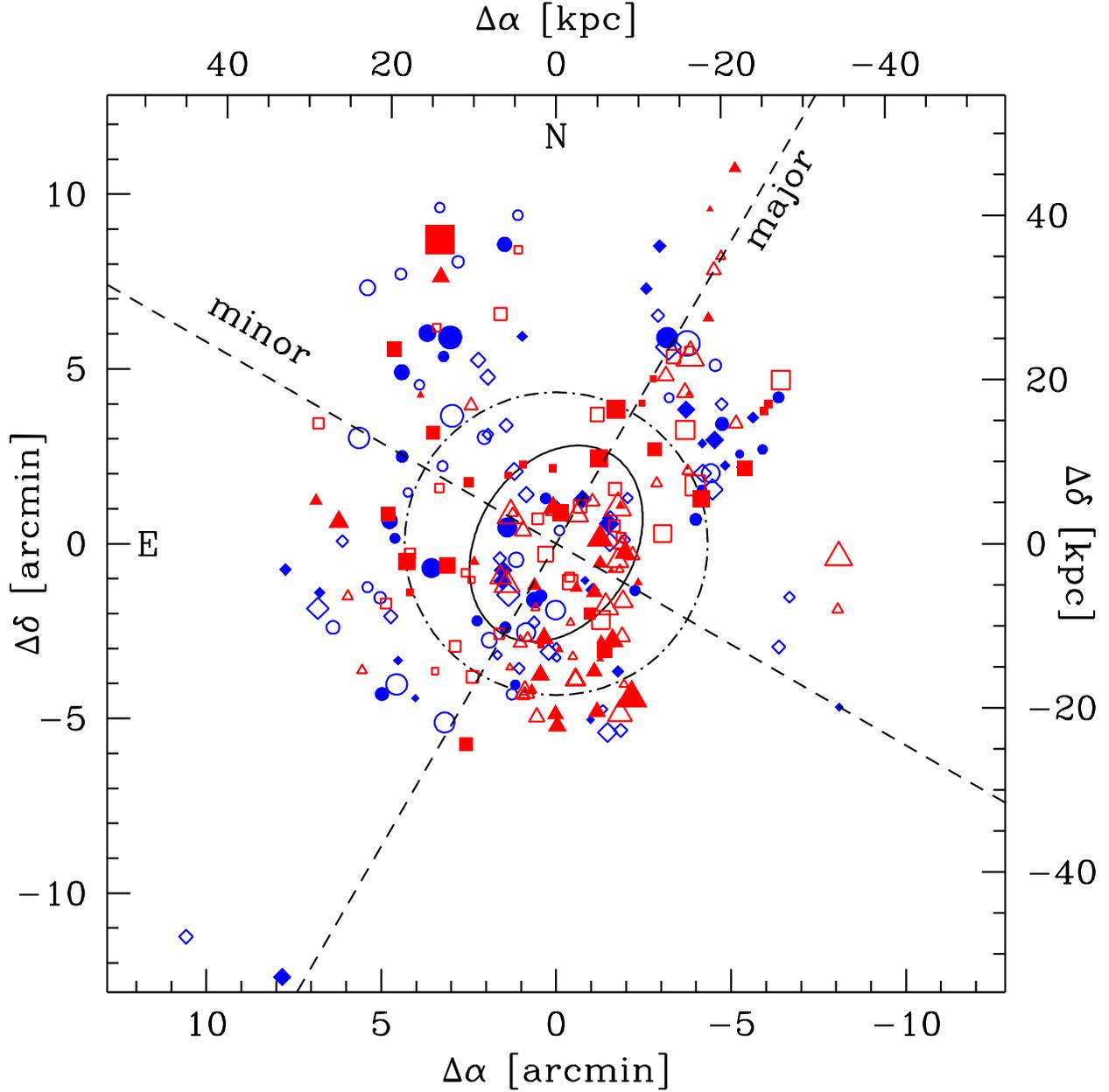} 
\caption{Spatial distribution of NGC 4636 GCs with measured velocities. 
Circles and diamonds represent the blue GCs measured in Paper I and \citet{sch06}, respectively.
Squares and triangles indicate the red GCs measured in Paper I and \citet{sch06}, respectively.
The GCs with velocities larger and smaller than
  the velocity ($\upsilon_{\rm gal}$ = 928 km s$^{-1}$) of the NGC 4636 nucleus are plotted
  by open symbols and filled symbols, respectively.
The symbol size is proportional to the velocity deviation.
The large solid-line ellipse represents a boundary for the standard diameter $D_{25}$ of NGC 4636 \citep{dev91}.
The photometric major and minor axes of NGC 4636 are represented by the dashed lines. 
The dot-dashed line circle marks a boundary ($R = 260^{\prime\prime}$) between the inner and outer region 
used for the analysis of the radial variation of kinematic properties.
\label{fig-vmap}}
\end{figure}

Of the entire spectroscopic sample of GC candidates 
we selected 238 member GCs of NGC 4636 in Paper I using radial velocities 
  ($300 \le v_p \le 1600~{\rm km~s}^{-1} $) and ($C-T_1$) colors ($0.9 \le C-T_1 < 2.1$). 
There are 108 blue GCs ($0.9\le(C-T_1)<1.55$) and 130 red GCs ($1.55\le(C-T_1)<2.1$) in the total sample.
In Figure \ref{fig-vmap}, 
  we show the spatial distribution of the objects in NGC 4636 with measured velocities.
It shows that the spatial segregation of high-velocity GCs (open symbols) and low-velocity GCs (filled symbols)
  is not clearly seen, which indicates that the rotation of the GC system is,
if any, weak 
  (to be discussed in \S \ref{rotation}).

\section{Results}\label{results}

We have investigated the kinematic properties of the GC system 
  using the master catalog of 238 GCs in NGC 4636 (Paper I):
  the rotation amplitude, the position angle of the rotation axis, 
  the mean line-of-sight velocity, the projected velocity dispersion, and the velocity ellipsoid.
\citet{cot01} and \citet{cot03} presented detailed analysis of the kinematics of the GC systems
  in M87 and M49, respectively.
We adopted the analysis method used by these studies for the following analysis, 
  as done for the M60 GC system \citep{hwa08}.

\subsection{Rotation of the GC System}\label{rotation}

First we derive the rotational property of the GC system in NGC 4636.
We made assumptions to derive the intrinsic rotational velocity field 
  from the radial velocity data:
  (a) the GC system is spherically symmetric with an intrinsic angular velocity field
  stratified on spheres, 
  and (b) that the GC rotation axis lies in the plane of the sky.
With these assumptions, it is expected that radial velocities of GCs depend,
if they follow any overall rotation, sinusoidally on the azimuthal angles.
For the NGC 4636 GC system, the assumption of spherical symmetry of the GC system is reasonable
  due to the modest projected ellipticity (effective ellipticity, $\epsilon_{\rm eff}=0.26$). 

Thus we can determine the amplitude and axis of the rotation, 
  fitting the observed line-of-sight velocities ($v_p$) of the GCs with the function,

\begin{equation}
v_p (\Theta) = v_{\rm sys} + (\Omega R) \sin(\Theta - \Theta_0)\ ,
\label{eq-rot}
\end{equation}

\noindent where $\Theta$ is the projected position angle of a GC relative to
  the galaxy center measured from north to east, 
  $\Theta_0$ is the projected position angle of the rotation axis of the GC system, 
  $R$ is the projected galactocentric distance,
  ($\Omega R$) is the rotation amplitude, and $v_{\rm sys}$ is the systemic velocity of the GC system.

\begin{figure}
\plotone{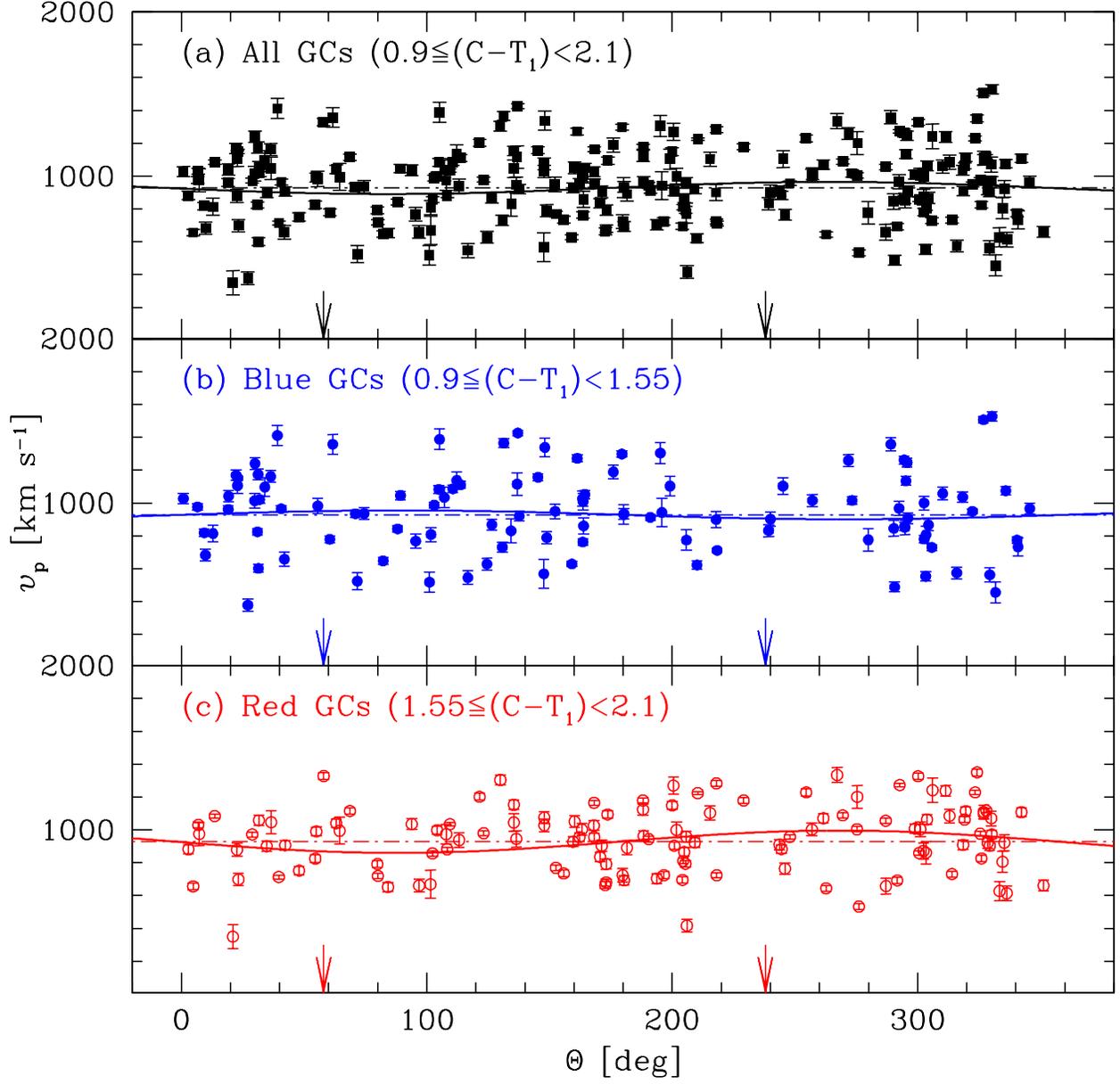} \caption{Radial velocities vs. position angles
  for all 238 GCs ({\it top}), 108 blue GCs ({\it middle}), and 130 red GCs ({\it bottom}). 
The solid curve represents the best fit rotation curve from Table \ref{tab-n4636kin}, 
  and the dot-dashed horizontal line indicates the velocity of the NGC 4636 nucleus. 
The photometric minor axis of NGC 4636 is represented by the vertical
  arrows ($\Theta = 58^{\circ}$ and $238^{\circ}$).
\label{fig-rotsub}}
\end{figure}

Figure \ref{fig-rotsub} displays the radial velocities of GCs with
  measured uncertainties as a function of position angle
  for all 238 GCs ({\it top}), 108 blue GCs ({\it middle}), and 130 red GCs ({\it bottom}).
We overlaid the best fit rotation curve of equation (\ref{eq-rot}) for each sample.
For fitting the data we used an error-weighted, nonlinear fit of equation (\ref{eq-rot})
  with $v_{\rm sys}$ as a fixed value of the velocity of the NGC 4636 nucleus 
  ($v_{\rm gal}=928\pm45$ km s$^{-1}$) rather than as a free parameter for a better fitting.
We estimated, using the biweight location of \citet{bee90}, 
  the systemic velocity of the GC
  system to be $v_{\rm sys}=949^{+13}_{-16}$ km s$^{-1}$ (see \S \ref{dispersion}),
  which agrees with the velocity of the NGC 4636 nucleus. 

The values of the rotation amplitudes ($\Omega R$) we derived  are
  $37^{+32}_{-30}$ km s$^{-1}$ for all the GCs,
  $27^{+34}_{-24}$ km s$^{-1}$ for the blue GCs, and
  $68^{+48}_{-35}$ km s$^{-1}$ for the red GCs.
Thus, all the GCs and the blue GCs show little rotation, 
  while the red GCs show a marginal hint of rotation.
Our results are consistent with those based on a smaller sample of 174 GCs in \citet{sch06} who derived
  $28\pm18$ km s$^{-1}$ for all the GCs,
  $11\pm27$ km s$^{-1}$ for the blue GCs, and
  $87\pm18$ km s$^{-1}$ for the red GCs.

The orientation of the rotation axis ($\Theta_0$)  is estimated to be
  ${174^\circ}^{+73}_{-48}$ for all the GCs,
  ${0^\circ}^{+146}_{-144}$ for the blue GCs, and
  ${178^\circ}^{+53}_{-34}$ for the red GCs.
The orientations of rotation axes for all subsamples appear to be similar,
  and they are closer to the photometric major axis, showing that the GCs rotate around
  the major axis ($\Theta_{phot} = 148^\circ$).
This result is in contrast with the result given by \citet{sch06} who derived 
  ${63^\circ}\pm43$ for all the GCs, and  ${60^\circ}\pm22$ for the red GCs,
  showing that the GCs rotate around the minor axis.
If we use only the data given by \citet{sch06}, we get similar values to those
given by \citet{sch06}. 
Therefore, the difference in the orientation between the two studies is due to the
  difference in the sample.

\begin{figure}
\plotone{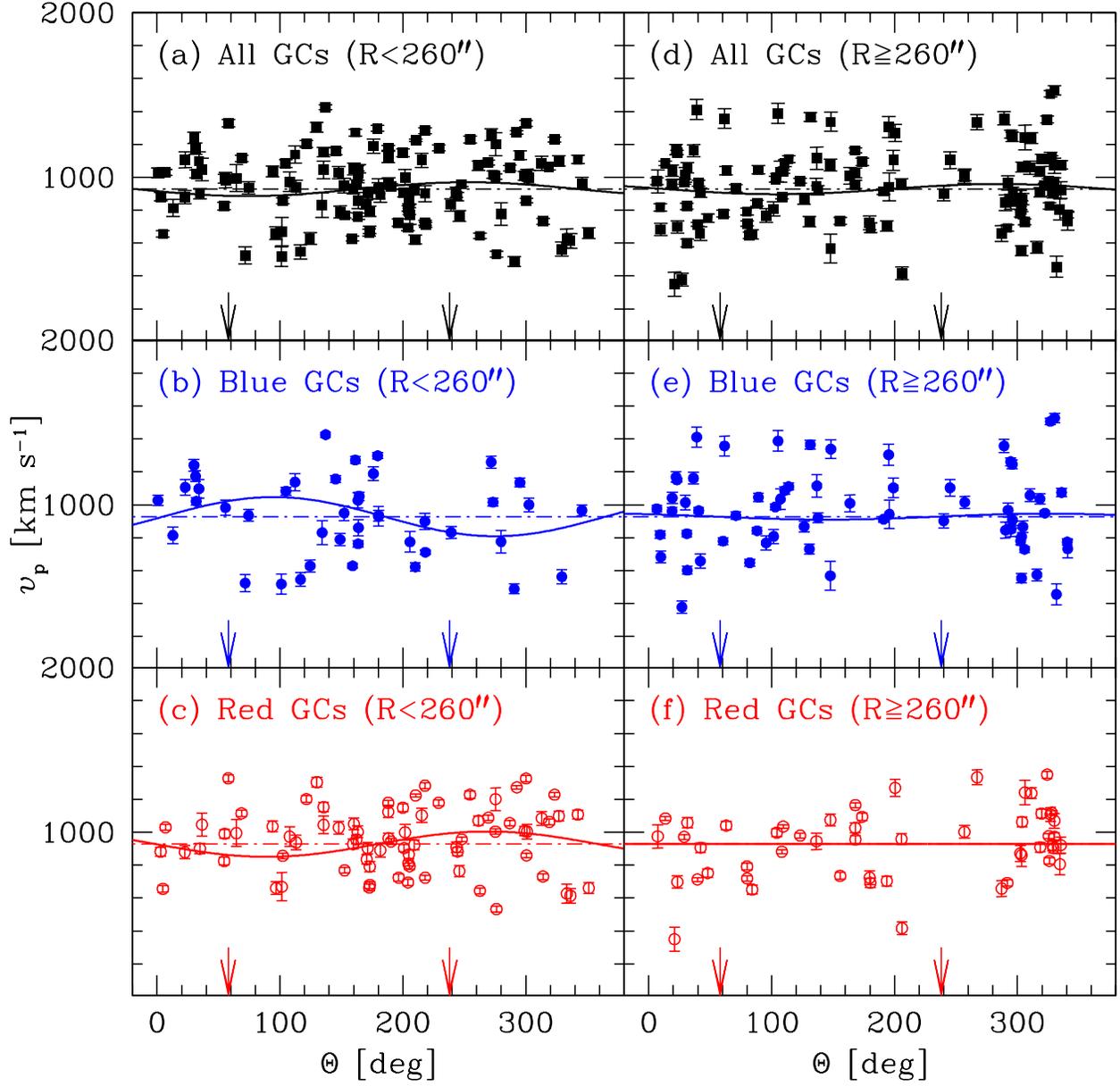} 
\caption{Radial velocities vs. position angles
for the GCs 
in the inner region ($23\arcsec \leq R < 260\arcsec$ , left panels) and 
the outer region ($260\arcsec \leq R < 926\arcsec$ , right panels). 
The best fit rotation curves for all the GCs, the blue GCs, and the red GCs 
  within each region are overlaid with solid lines. 
The dot-dashed horizontal line indicates the 
  velocity of the NGC 4636  nucleus, and the vertical arrows
  mark the position angle of photometric minor axis of NGC 4636.
\label{fig-rotreg}}
\end{figure}

In Figure \ref{fig-rotreg}, we present the rotation of the GC
  system for the samples of different radial bins in order to investigate
  any radial variation of rotational properties.
We divided the GCs into two groups:
 the GCs in the inner region at $23\arcsec \leq R < 260\arcsec$ and 
  those in the outer region at  $260\arcsec \leq R < 926\arcsec$. 
Each of the sub regions includes a similar number of GCs (120 and 118, respectively).
We applied the same fitting procedure as used for the results in Figure \ref{fig-rotsub} 
  for all the GCs, the blue GCs, and the red GCs in each radial bin.

Interestingly both the blue GCs and red GCs show some rotation around the major axis
  in the inner region ($119^{+72}_{-62}$ km s$^{-1}$, and  $77^{+44}_{-47}$ km s$^{-1}$, respectively),
  while both subsamples show little rotation in the outer region.
The rotation of the red GCs seen for the entire radial range is found to be 
  mainly due to the red GCs in the inner region.
Little rotation is seen for all the GCs even in the inner region, 
  although both the blue GCs and red GCs show measurable rotation in the inner region. 
This is because the blue GCs and red GCs rotate in the opposite direction to each other,
canceling the rotational effect for all the GCs.

\subsection{Velocity Dispersion of the GC System}\label{dispersion}

We summarize the kinematics of the NGC 4636 GC system derived in this study in Table \ref{tab-n4636kin}. 
Several kinematic parameters for all the GCs, the blue GCs, and the red GCs are presented 
 for the entire region, 
 the inner region, 
 and the outer region. 
The column (1) defines the range of the projected radial distance from the center of NGC 4636
  for each region in arcsec, 
  and the column (2) gives the mean value of  the radial distance in arcsec.
The number of GCs in each region is shown in the column (3). 
The columns (4) and (5) represent the mean line-of-sight velocity (the biweight location of \citealt{bee90})
  and the velocity dispersion about this mean velocity (the biweight scale of \citealt{bee90}), respectively. 
The position angle of the rotation axis and the rotation amplitude estimated using
  equation (\ref{eq-rot}) in each region are given in the columns (6) and (7), respectively.
The column (8) gives the velocity dispersion about the best fit rotation curve ($\sigma_{p,r}$). 
The column (9) gives the absolute value
  of the ratio of the rotation amplitude to the velocity dispersion about the best fit rotation curve. 
The uncertainties of these values represent $68\%$ $(1\sigma)$ confidence intervals that are
  determined from the numerical bootstrap procedure following the method of \citet{cot01}.

The velocity dispersion for all 238 GCs is estimated to be $225^{+12}_{-9}$ km s$^{-1}$. 
This result is consistent, within the uncertainty, with the value given by \citet{sch06}
  who used 174 GCs,  $203\pm11$ km s$^{-1}$. 
In addition, it is found that the velocity dispersion about the mean velocity of the GC system
  for the blue GCs  ($\sigma_p = 251^{+18}_{-12}$ km s$^{-1}$) is 46 km s$^{-1}$~ larger than
  that for the red GCs ($\sigma_p = 205^{+11}_{-13}$ km s$^{-1}$).
We found a similar result for the velocity dispersion about the best fit rotation curve:
    $\sigma_{p,r}=252^{+15}_{-15}$ km s$^{-1}$ for the blue GCs, 
  and $\sigma_{p,r}=203^{+12}_{-13}$ km s$^{-1}$ for the red GCs.
Our result for the red GCs is similar to the value given by \citet{sch06}, 
  $\sigma_p = 199\pm17$ km s$^{-1}$, within the uncertainty. 
However, our value for the blue GC is approximately 50 km s$^{-1}$ larger than
  that given by \citet{sch06}, $\sigma_p = 202\pm15$ km s$^{-1}$.

\begin{figure}
\plotone{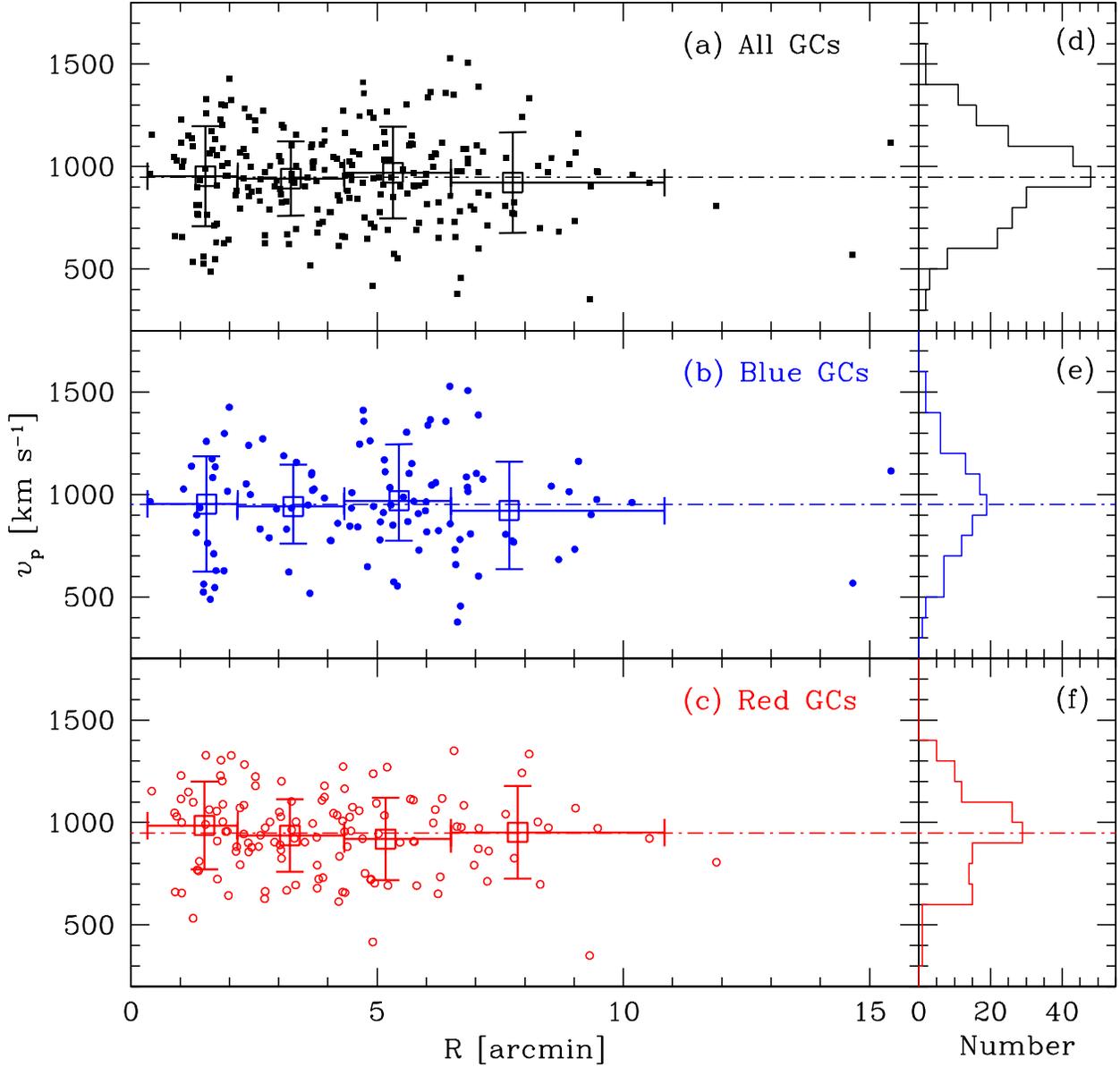} \caption{Radial velocities vs. projected
  galactocentric distances for all the GCs ({\it top}), the blue GCs ({\it middle}),
  and the red GCs ({\it bottom}). 
Large open squares represent the mean radial velocities of GCs in the radial bins that
  are represented by long horizontal error bars. 
Their vertical error bars denote the velocity dispersions of GCs in the radial bins. 
The histograms in the right panels represent the velocity distribution of each sample.
The dot-dashed horizontal line indicates the systemic velocity of NGC 4636. 
\label{fig-vel}}
\end{figure}

Figure \ref{fig-vel} displays the radial velocities of GCs with
measured uncertainties against projected galactocentric distances. 
The mean radial velocities in four radial bins are overlaid by squares with long horizontal error bar.
The velocity dispersion about the mean velocity in each bin is also represented
  by a vertical error bar. 
The mean velocities of all samples agree well with the systemic velocity of NGC 4636.  

\begin{figure}
\plotone{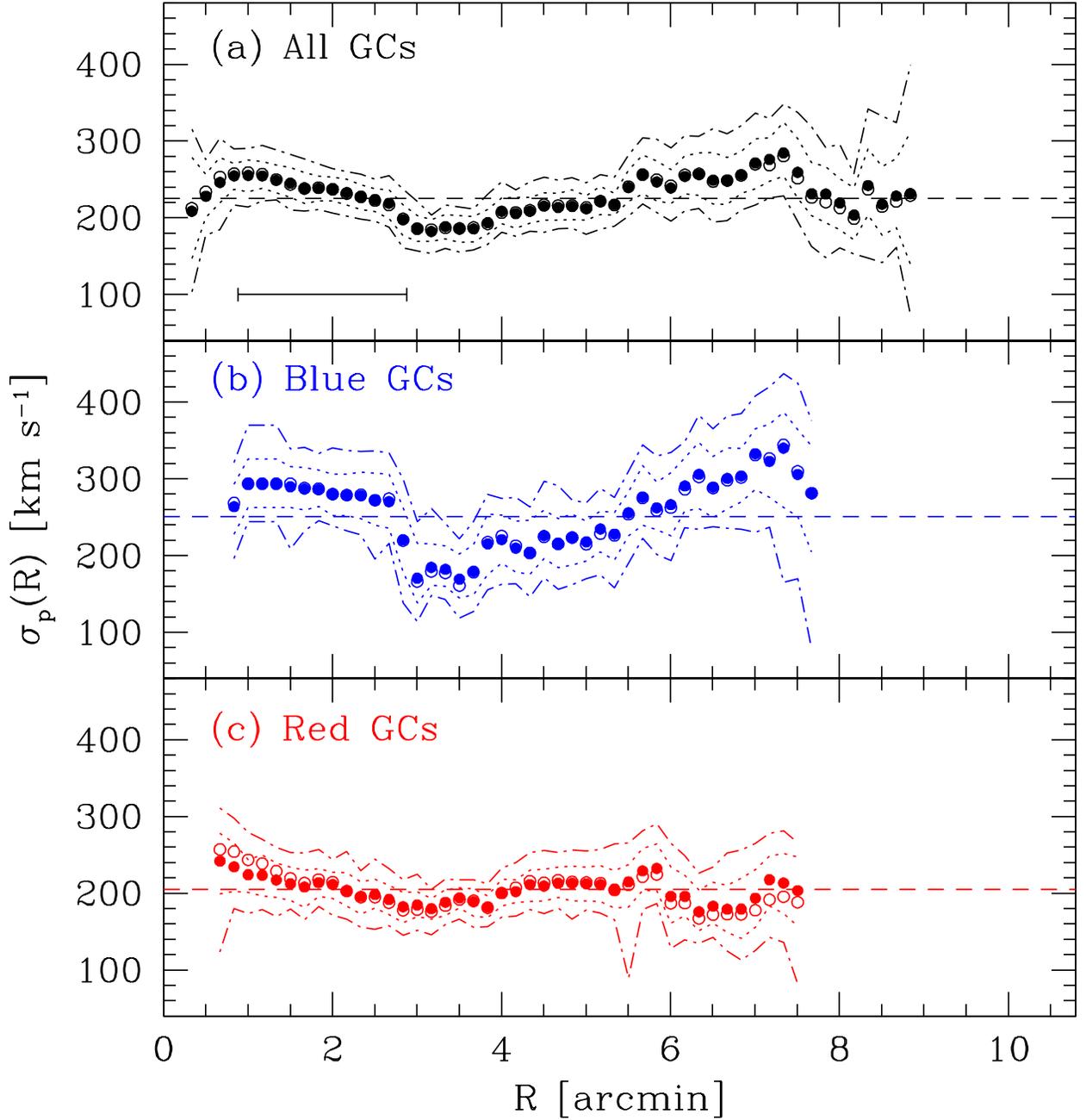} 
\caption{Radial profiles of velocity dispersion
  for all the GCs ({\it top}), the blue GCs ({\it middle}), and the red GCs ({\it bottom}). 
Filled circles represent the velocity dispersion about the mean GC velocity
($\sigma_p$) at each point, while open circles the velocity
dispersion about the best fit rotation curve ($\sigma_{p,r}$) at
the same point. The dispersion is calculated using the GCs within
a moving radial bin (width of $2\arcmin\simeq 8.52$ kpc) that is
represented by a horizontal error bar in top panel. The dotted and
dot-dashed lines denote 68\% and 95\% confidence intervals on the
calculation of velocity dispersion, respectively. The dashed
horizontal line indicates the global value of velocity dispersion
of GCs in each panel. 
\label{fig-disp}}
\end{figure}

To investigate the radial variation of velocity dispersion in detail,
  we present a smoothed radial profile of velocity
  dispersion about the mean radial velocity (filled symbols) and about the
  best fit rotation curve (open symbols) in Figure~\ref{fig-disp}.
We calculate the velocity dispersion of the GCs lying within a radial bin with fixed  width,
  $\Delta R = 120\arcsec \simeq 8.52$ kpc as increasing the
  bin center by a fixed step width, $\delta R = 10\arcsec\simeq 0.71$ kpc.
We set the radial width and the step width so
  that the number of GCs per bin exceeds 10, and the calculation stops
  when the number of GCs in a bin is smaller than 10.

The velocity dispersion for all the GCs varies more or less (about 50 km s$^{-1}$) 
  around the mean value for the range of radius. 
However, the velocity dispersion for the blue GCs shows an abrupt decrease from $\approx300$ km s$^{-1}$
  to $\approx 150$  km s$^{-1}$ at $R \approx 3\arcmin $, 
  and then keeps increasing with increasing radius to $\approx 300$ km s$^{-1}$ at $R \approx 7\arcmin $. 
On the other hand, the velocity dispersion for the red GCs stays almost constant 
at $\approx 200$   km s$^{-1}$  from the center out to $R \approx 6\arcmin $, 
shows  a slight decrease of $\approx 50$ km s$^{-1}$ for $6\arcmin < R < 7\arcmin $, 
and increases slightly at  $ R \approx 7\arcmin $. 
 The velocity dispersions about the best fit rotation curves of all three samples
are not much different from those about the mean radial velocities (also see Table \ref{tab-n4636kin}).

\subsection{Velocity Anisotropy of the Globular Cluster System}

Assuming that the NGC 4636 GC system is spherically symmetric, 
  we can apply the Jeans equation in the absence of rotation to the
  dynamical analysis of the GC system. 
The spherical Jeans equation is represented by

\begin{equation}
{d\over{dr}}\, n_{\rm cl}(r) \sigma_r^2(r) + 
  {{2\,\beta_{\rm cl}(r)}\over{r}}\, n_{\rm cl}(r) \sigma_r^2(r) = - 
  n_{\rm cl}(r)\,{{G M_{\rm tot}(r)}\over{r^2}}\ , 
\label{eq-jeans}
\end{equation}

\noindent where $r$ is a three dimensional radial distance from the galactic center,
  $n_{\rm cl}(r)$ is a three dimensional density profile of the GC system,
  $\sigma_r(r)$ is a radial component of velocity dispersion,
  $\beta_{\rm cl}(r)\equiv1-\sigma_\theta^2(r)/\sigma_r^2(r)$ is a velocity anisotropy,
  $G$ is the gravitational constant,
  and $M_{\rm tot}(r)$ is a total gravitating mass contained
  within a sphere of radius $r$ (e.g., \citealt{bt87}).
$\sigma_\theta(r)$ is a tangential component of velocity dispersion that is
  equal to an azimuthal component of the velocity dispersion, $\sigma_\phi(r)$,
  in the absence of rotation of the GC system.

There are two approaches to the dynamical analysis of the GC system:
     a) to derive the mass profiles of the host galaxy for given orbits of the GCs, 
  or b) to determine the orbital properties of the GCs for a given mass profile of the host galaxy.
Several studies on the dynamics of the GC system have focused on
  determining the gravitational mass, $M_{\rm tot}(r)$, using the Jeans
  equation by assuming a simple isotropic orbit with $\beta_{\rm cl}(r)=0$ 
  (e.g., \citealt{coh97,min98,zep00,sch06}).

However, with an aid of an independent determination of the mass profile of an
  elliptical galaxy using X-ray data (e.g., \citealt{bri97,mat98,loe03} for NGC 4636),
  the velocity anisotropy itself can be investigated  (e.g., \citealt{rom01,cot01,cot03, hwa08}).
Following the analysis of the M87 GC system by \citet{cot01}, the M49 GC system by \citet{cot03}, 
  and the M60 GC system by \citet{hwa08}, 
  we first derive the three dimensional density profile of the GC system, $n_{\rm cl}(r)$ 
  and the total mass profile, $M_{\rm tot}(r)$.
Comparing the velocity dispersion profile (VDP) calculated from the Jeans equation
  with the observed VDP, $\sigma_p(R)$, we determine
  the velocity anisotropy of the NGC 4636 GC system.
  
\subsubsection{Density Profiles for the GC system}

Here we derive the three dimensional density profiles from the surface number density
  of NGC 4636 GCs for two models: the NFW profile \citep{nfw97} and the Dehnen profile \citep{deh93}.
We used the surface number density profiles of NGC 4636 GCs given in \citet{par09b}.
They derived the surface density profile
  of NGC 4636 GCs by combining the HST/WFPC2 archive data
  for the inner region at $R<1.5\arcmin$,
  and the KPNO data for the outer region at $R>1.5\arcmin$.
They adopted the background levels of the mean surface number density 
from \citet{dir05} at $R>13.5\arcmin$:
  $2.33\pm 0.07$ per square arcmin for all the GCs,
  $1.88\pm 0.06$ per square arcmin for the blue GCs,
  and  $0.76\pm 0.04$ per square arcmin for the red GCs.
Then they subtracted these background values from the original number counts
  to produce the radial profiles of the net
  surface number density of GCs.
Since they selected GCs that are brighter than $T_1 \approx 23.0$ mag, 
  it is needed to correct the surface number density
  profile in order to account for the uncounted GCs due to the
  limiting magnitude.
To calculate the correction factor, the equation (11) in \citet{mcl99} with $V_{lim,1}=\infty$ was
  used on the assumption that the GC luminosity function of NGC 4636 has
  a Gaussian shape with a peak at $T_1 \approx 23.31$ mag and a
  dispersion $\sigma = 1.03 $ mag \citep{dir05}.
It is found that the surface number density of the bright NGC 4636 GCs in \citet{par09b} 
should be multiplied by 2.62  to derive the total surface number density.

\begin{figure}
\plotone{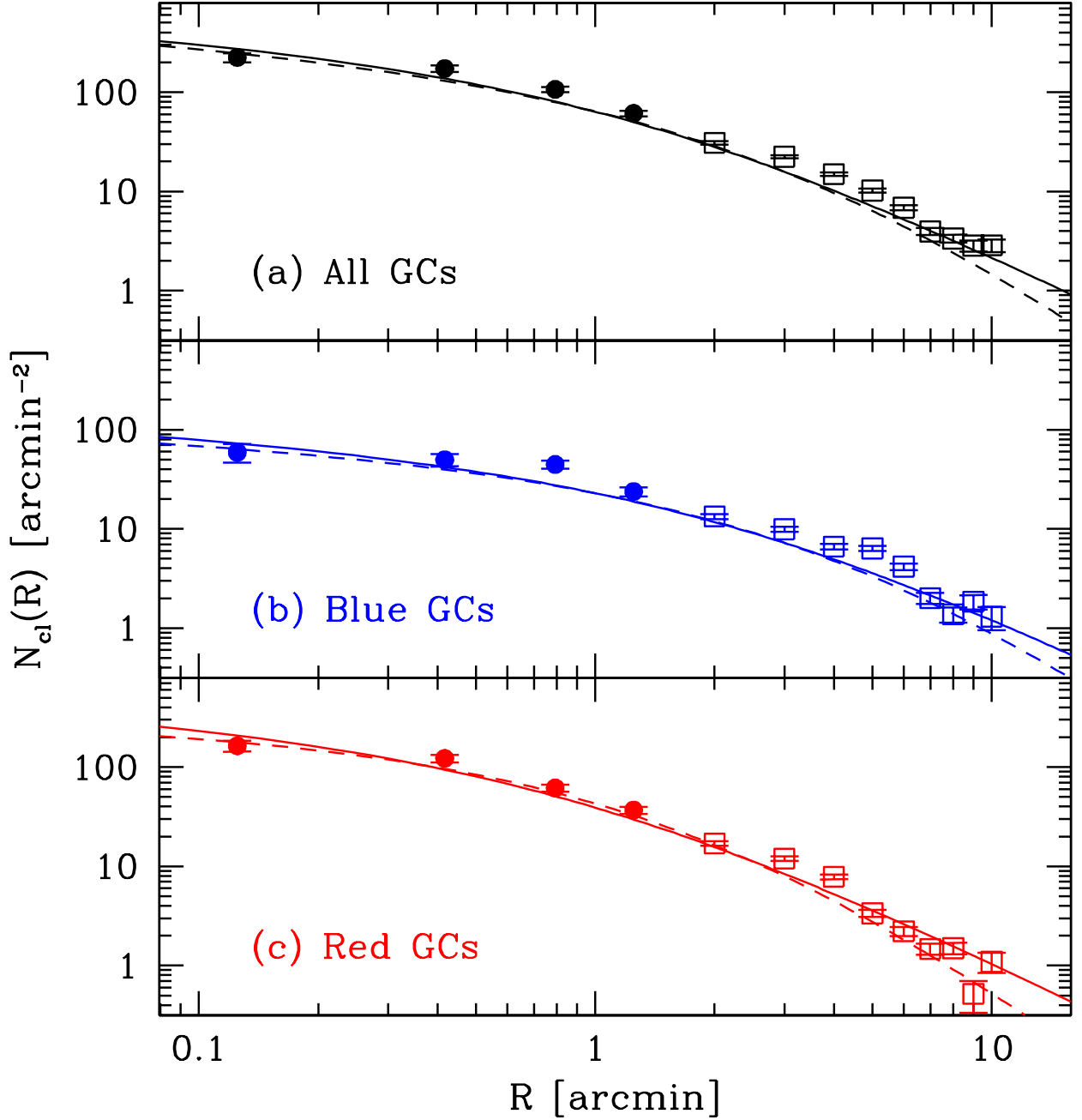} 
\caption{Projected number density profiles for all the GC candidates ({\it top}), 
  the blue GC candidates ({\it middle}), and the red GC candidates ({\it bottom})
  in the photometric catalog.
Filled circles represent the GC candidates from the HST/WFPC2 images,
  while open squares the GC candidates from the KPNO $CT_1$ images \citep{par09b}. 
The solid line and the dashed line in each panel indicate the projected best fits using
  the NFW density profile and the Dehnen density profile, respectively, for each sample. 
\label{fig-numden}}
\end{figure}

We display the total surface number density profiles, $N_{\rm cl}(R)$, for
  the combined, blue and red GCs in Figure \ref{fig-numden}.
We fit the surface number density profile with the projection of the NFW profile, 
  $n_{\rm cl} (r)=n_0(r/b)^{-1}(1+r/b)^{-2}$ and 
  with the projection of the Dehnen profile,
  $n_{\rm cl} (r)=n_0(r/a)^{-\gamma}(1+r/a)^{\gamma-4}$.
The surface number density profile, $N_{\rm cl}(R)$, is
  derived from the integration of the three dimensional density profile $n_{\rm cl}(r)$ as follows:
\begin{equation}
N_{\rm cl}(R) = 2\int_{R}^{\infty} n_{\rm cl}(r) {{r\,dr}\over{\sqrt{r^2-R^2}}}\ . 
\label{eq-ncl}
\end{equation}

The solid and long dashed lines represent the projected best fit curves
  of the NFW profile and of the Dehnen profile, respectively.
The fitting results for the combined (C) GCs, blue (B) GCs, and red (R) GCs
  are summarized as follows:

\begin{equation}
\begin{array}{rcl}
n_{\rm cl}^{\rm C}(r) & = &
  0.57\,{\rm kpc}^{-3}(r/6.17\,{\rm kpc})^{-1}(1+r/6.17\,{\rm kpc})^{-2} \\
n_{\rm cl}^{\rm B}(r) & = &
  0.07\,{\rm kpc}^{-3}(r/10.71\,{\rm kpc})^{-1}(1+r/10.71\,{\rm kpc})^{-2} \\
n_{\rm cl}^{\rm R}(r) & = &
  0.71\,{\rm kpc}^{-3}(r/4.39\,{\rm kpc})^{-1}(1+r/4.39\,{\rm kpc})^{-2},  \\
\end{array}
\label{eq-nfw}
\end{equation}
for the NFW profile, and
\begin{equation}
\begin{array}{rcl}
n_{\rm cl}^{\rm C}(r) & = &
  0.14\,{\rm kpc}^{-3}(r/15.87\,{\rm kpc})^{-1.08}(1+r/15.87\,{\rm kpc})^{-2.91} \\
n_{\rm cl}^{\rm B}(r) & = &
  0.03\,{\rm kpc}^{-3}(r/21.14\,{\rm kpc})^{-0.94}(1+r/21.14\,{\rm kpc})^{-3.06} \\
n_{\rm cl}^{\rm R}(r) & = &
  0.47\,{\rm kpc}^{-3}(r/8.40\,{\rm kpc})^{-0.70}(1+r/8.40\,{\rm kpc})^{-3.30},  \\
\end{array}
\label{eq-dh}
\end{equation}
for the Dehnen profile.

It is found that the scale length $b$ of the red GCs in the NFW profile  is 
more than twice smaller
  than that of the blue GCs. This  shows that the red GCs are more
  concentrated toward the galaxy center than the blue GCs. 

\subsubsection{An Extended Dark Matter Halo in NGC 4636}\label{dark}

We investigate the existence of an extended dark matter halo 
  comparing the velocity dispersion profile expected from the stellar mass profile
  with the measured profiles of stellar and GC velocity dispersions for NGC 4636. 
We first derive the stellar mass profile from the surface brightness profile. 
Then we obtain the velocity dispersion profile expected from the stellar mass profile, 
  which will be compared with the measured dispersion profiles for the GCs.

\begin{figure}
\plotone{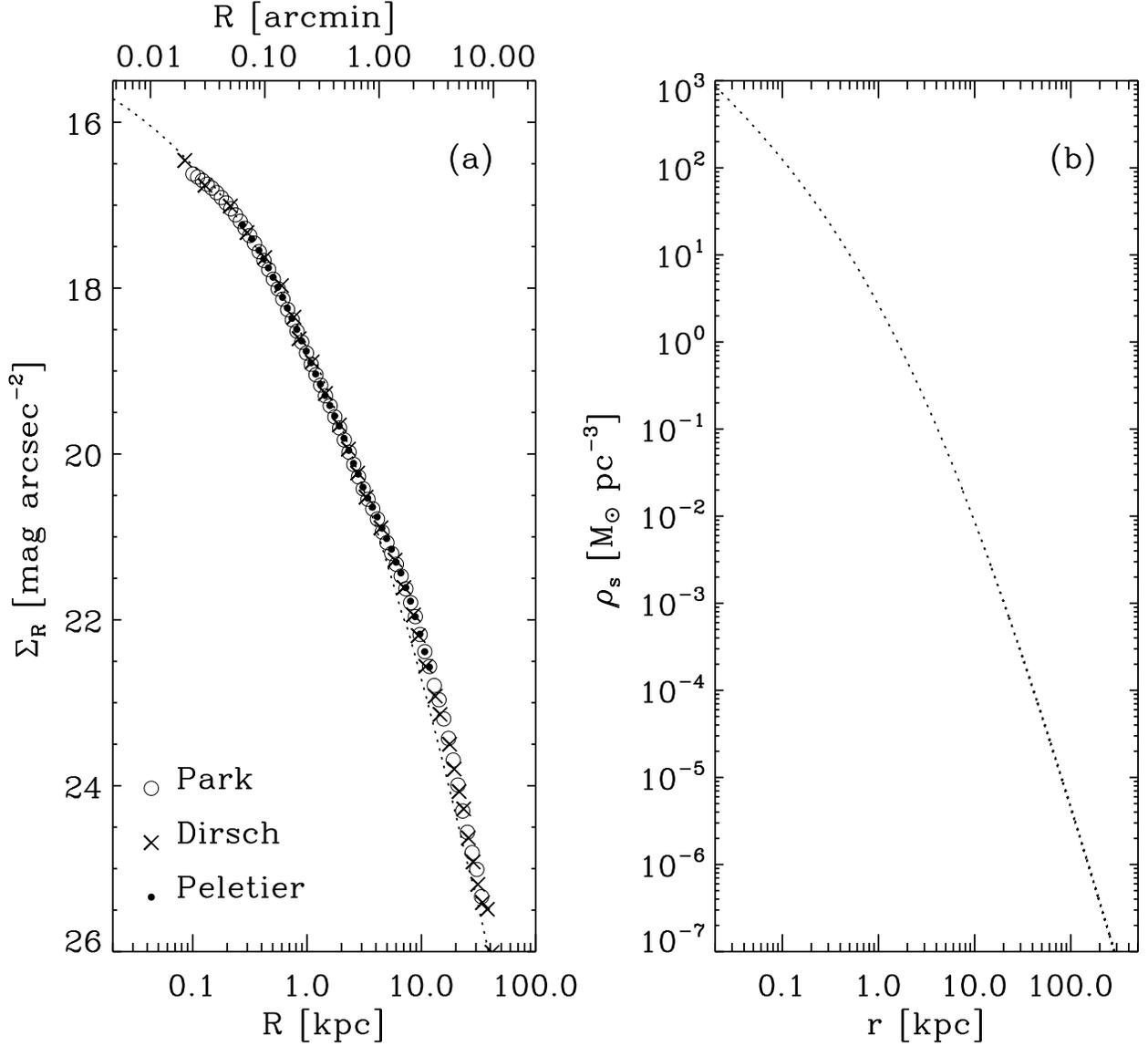} 
\caption{ 
{\it Left}: $R$-band surface photometry of NGC 4636 derived from KPNO images 
  \citep[open circles]{par09b}, \citet[crosses]{dir05}, and \citet[filled circles]{pel90}. 
The dotted line indicates a projected best fit using eq. (\ref{eq-lumden}).
{\it Right}: Three dimensional stellar mass density profile
  using the best fit model in the left panel with
  a constant $R$-band mass-to-light ratio of $\Upsilon_0=9.0~M_\odot L^{-1}_{R,\odot}$.
\label{fig-surfphot}}
\end{figure}

In the left panel of Figure \ref{fig-surfphot},
  we plot the surface brightness profile of NGC 4636
  derived from our KPNO $T_1$-band images \citep{par09b}
  compared to those in \citet{pel90, dir05} for the $R$-band photometry.
We converted $T_1$ photometry of \citet{par09b} to Cousins $R$-band photometry
  using the relation given by \citet{gei96}.
It is seen that the profiles from this study, \citet{pel90}, and \citet{dir05} 
  agree well over the radius.

We fit the surface brightness profile derived from the KPNO images \citep{par09b}
  with the projection of three dimensional luminosity density profile used in \citet{cot03},
  which is represented by,
\begin{equation}
j(r) = {{(3-\gamma)(7-2\gamma)}\over{4}}\, {L_{\rm tot}\over{\pi a^3}}\, 
  \left({r\over{a}}\right)^{-\gamma}\,
  \left[1+\Bigl({r\over{a}}\Bigr)^{1/2} \right]^{2(\gamma-4)}\ .
\label{eq-lumden}
\end{equation}

We obtain, from the fit, the values for the parameters,
  $\gamma=0.84$, $L_{\rm tot}=5.42\times10^{10}~L_{R,\odot}$, and $a=2.61$ kpc,
  and overlay the projected best fit curve  in Figure \ref{fig-surfphot}.
We also derive a value for the effective radius of 
  $R_{\rm eff}=2.\arcmin58 \simeq 10.99$ kpc, which is 
  larger than that from a fit ($R_{\rm eff}=1.\arcmin49 \simeq 6.347$
  kpc at $T_1$-band) using a de Vaucouleurs law in \citet{par09b}.

In the right panel of Figure \ref{fig-surfphot},
  we show a three dimensional stellar mass density profile, $\rho_s(r)=\Upsilon_0 j(r)$,
  with $R$-band mass-to-light ratio $\Upsilon_0=9.0~M_\odot L^{-1}_{R,\odot}$
  (to be discussed later in this section).
From the three dimensional stellar mass density profile 
we derive a stellar mass profile of NGC 4636 that  is represented by
\begin{eqnarray}
M_{\rm s}(r) & = &\int_{0}^{r}{4{\pi}x^2\,\rho_s(r)}dx  = \Upsilon_0\int_{0}^{r}{4{\pi}x^2\,j(x)}dx \nonumber \\
 & = & {\Upsilon_{0}}L_{\rm tot}\left[{(r/a)^{1/2}
    \over 1+(r/a)^{1/2}}\right]^{2(3-\gamma)}\left[{(7-2\gamma)+(r/a)^{1/2}
    \over 1+(r/a)^{1/2}}\right] \ .
\label{eq-starmass}
\end{eqnarray}

We used this stellar mass profile to determine the velocity
  anisotropy for the NGC 4636 stellar system and to test the existence of
  an extended dark matter halo.
If we take $M_{\rm tot}(r)=M_{\rm s}(r)$ and substitute $n_{\rm cl}(r)$ by $\rho_s(r)\propto j(r)$,
  then we can compute the radial component of velocity dispersion profiles (VDPs) of the stars
  through the Jeans equation
  by assuming the values for $R$-band mass-to-light ratios ($\Upsilon_0$) and
  velocity anisotropies of the stellar system [$\beta_{\rm s}(r)$].
We therefore obtain the projected VDPs for the stellar system
  from the radial component of VDPs through the equation (11) in Section 3.3.4.

\begin{figure}
\plotone{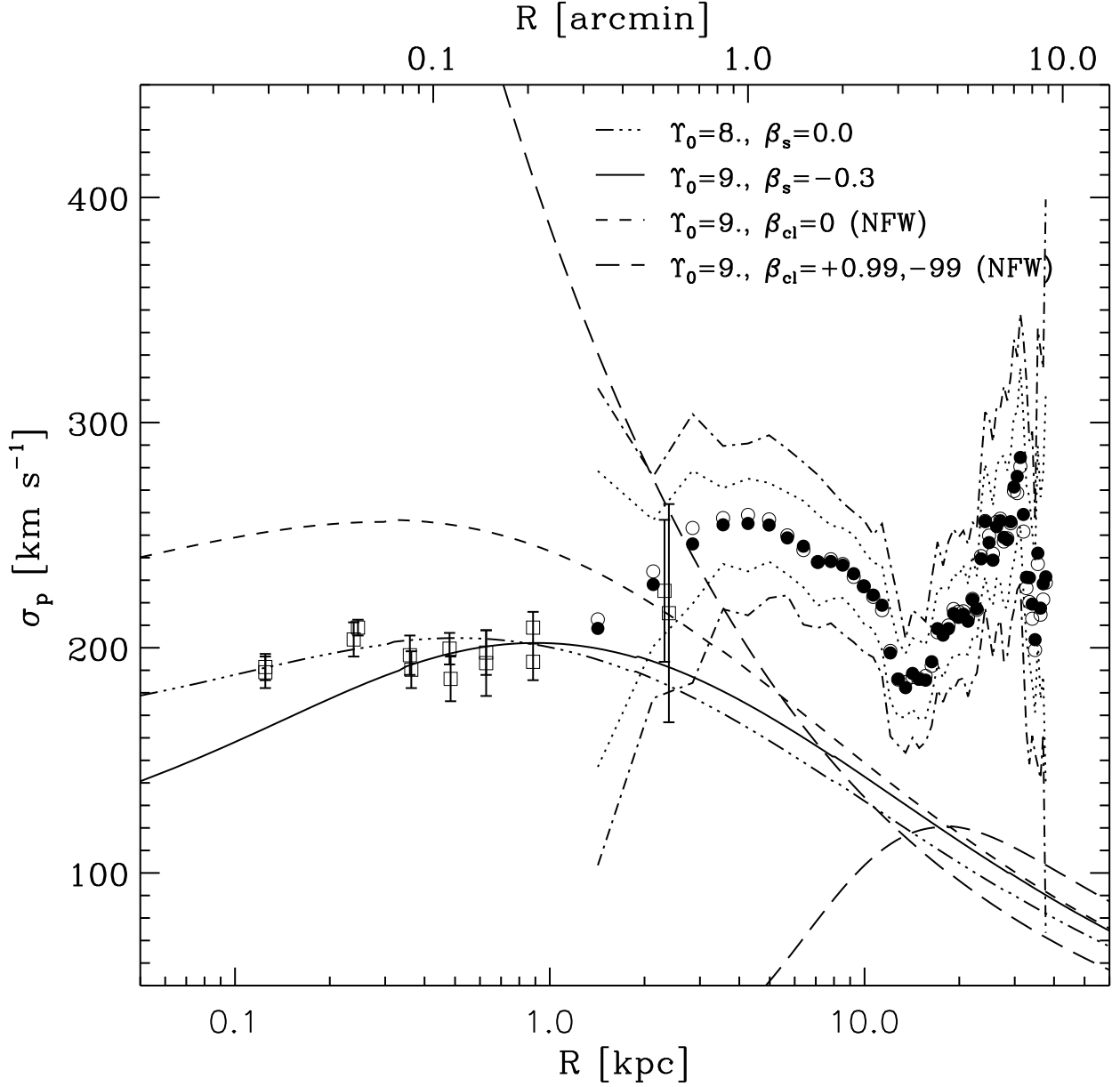} 
\caption{VDPs for the stars and the GCs.
Stellar VDPs are from \citet[open squares]{ben94}, and the GC VDPs are shown by filled and open circles 
  with associated dotted and dot-dashed lines (from Fig. \ref{fig-disp}).
The solid line represents the stellar VDP calculated using the stellar mass model in Fig. \ref{fig-surfphot} with
  a constant stellar mass-to-light ratio of $\Upsilon_0=9.0~M_\odot L^{-1}_{R,\odot}$
  and a stellar velocity anisotropy of $\beta_{\rm s}=-0.3$.
The double-dot-dashed curve shows  the stellar VDP with $\Upsilon_0=8.0~M_\odot L^{-1}_{R,\odot}$
  and $\beta_{\rm s}=0.0$.
Other lines represent the VDPs calculated using the same stellar mass model as above
  with NFW GC density profiles and velocity anisotropies
  of $\beta_{\rm cl}=0.0$ (short dashed line), $+0.99, -99$ (long dashed lines). 
\label{fig-stardisp}}
\end{figure}

In Figure \ref{fig-stardisp}, we plot the stellar VDP for the inner region ($R<34\arcsec$)
 of NGC 4636 given in \citet{ben94} and the GC VDP derived in this study.
The stellar VDP is almost constant around 200 km s$^{-1}$ in the inner region,
 and is smoothly connected to the GC VDP at $R\approx 2$ kpc that is increasing as the radius increases.
Note that the GC VDP varies much more than the stellar VDP.
In Figure \ref{fig-stardisp}, we also show the projected VDPs calculated with 
  $\Upsilon_0=8.0~M_\odot L^{-1}_{R,\odot}$, $\beta_{\rm s}(r)=0.0$ (isotropic),
  and $\Upsilon_0=9.0~M_\odot L^{-1}_{R,\odot}$, $\beta_{\rm s}(r)=-0.3$ (tangentially biased),
  which are the best fit curves for the stellar kinematic data of \citet{ben94} at $R<1$ kpc.
These values for the mass-to-light ratio and velocity anisotropy for the stellar system 
are  similarly found in the ``best-fitting'' halo model using $B$-band photometry of NGC 4636 
by \citet{kro00}. 

For the comparison, we also present the projected VDPs calculated
  using the same stellar mass profile as above,
  but for the GC number density profile $n_{\rm cl}(r)$
  and for $\beta_{\rm cl}(r)=+$0.99 (radially biased; upper long dashed line),
  --99 (tangentially biased; lower long dashed line),
  0.0 (isotropic; short dashed line). 
Interestingly, none of these models can account for
  the observed VDPs for the GCs at $R>3$ kpc, indicating that
  mass-to-light ratio is not constant over the galactocentric distance, 
  but should increase as the distance increases.
This demonstrates that there exists an extended dark mater halo in the outer region of NGC 4636.

\subsubsection{X-ray Mass Profiles}

\begin{figure}
\plotone{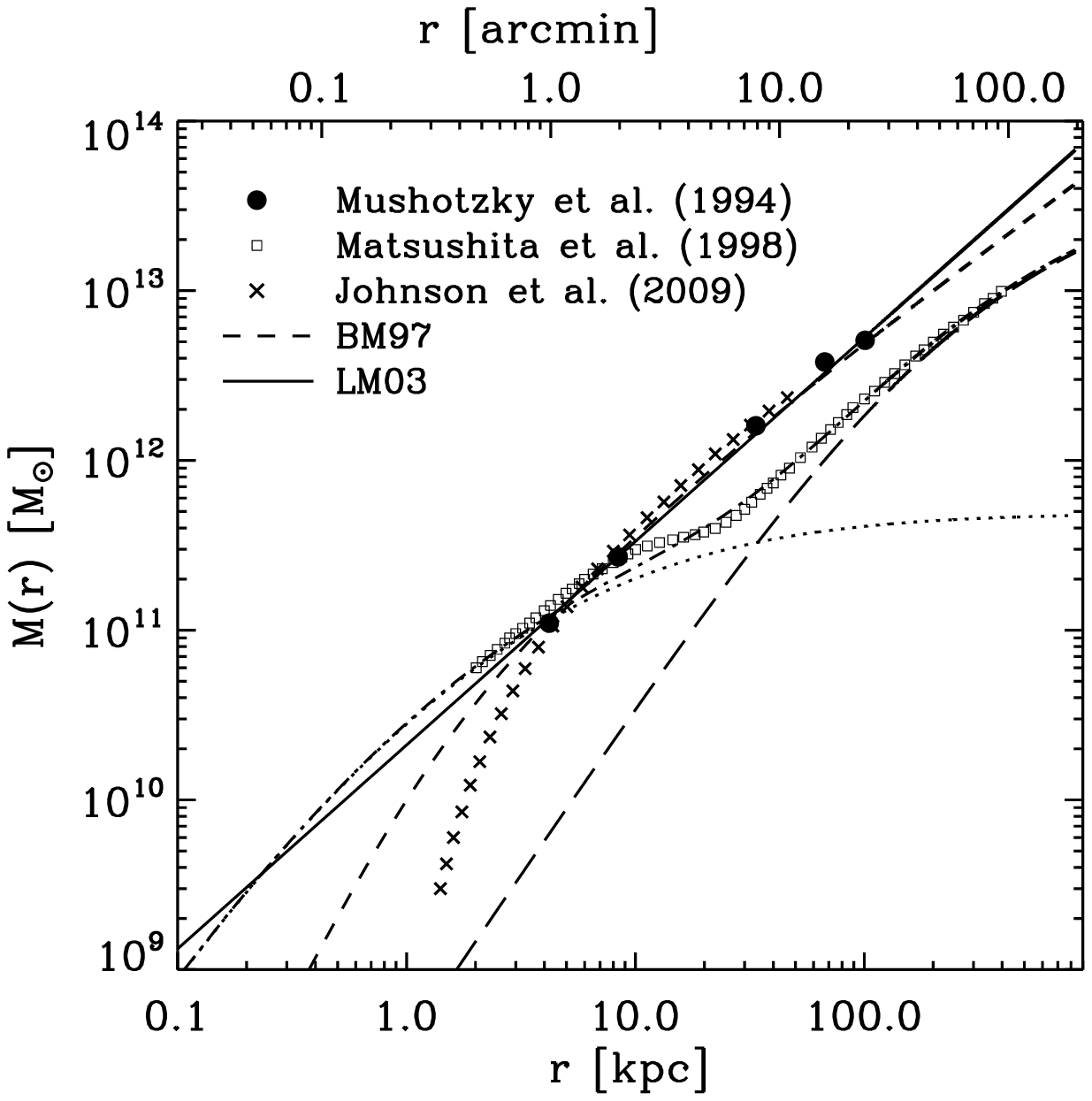} 
\caption{Total mass profiles of NGC 4636 derived from X-ray data.
The solid line, short-dashed line, filled circles, open squares, and crosses represent the  mass profiles derived 
  by Loewenstein \& Mushotzky (2003, LM03), Brighenti \& Mathews (1997, BM97), 
  \citet{mus94}, \citet{mat98}, and \citet{joh09}, respectively.
The dotted line represents the stellar mass profile derived
  using a constant $R$-band mass-to-light ratio of $\Upsilon_0=9.0~M_\odot L^{-1}_{R,\odot}$ from the result of \citet{mat98}.
The dot-dashed line and the long-dashed line represent, respectively, the profiles of the total mass
  and the dark matter derived from the result of \citet{mat98}.
\label{fig-mass}}
\end{figure}

There are several studies that presented the X-ray mass profiles of NGC 4636 
(e.g., \citealt{mus94,bri97,mat98,loe03,joh09}), as displayed in Figure \ref{fig-mass}.
\citet{bri97} derived a mass profile of NGC 4636 
  using {\it Einstein} HRI observational data of \citet{tri86} and 
  {\it ROSAT} PSPC observational data of \citet{tri94}.
They presented a functional form of the mass profile,
  which is used for the analysis of velocity anisotropy in this study.
Their mass profile 
 agrees well with that given by \citet{mus94} who used {\it ASCA} SIS data.

\citet{loe02,loe03} determined a mass profile of NGC 4636
  using {\it Chandra} and {\it XMM-Newton} X-ray data.
They showed that the total mass of NGC 4636 increases as $r^{1.2}$ to a good approximation 
  from 0.7 to 35 kpc with $\sim 1.5\times 10^{12} M_{\sun}$ at the outermost point.
Recently \citet{joh09} derived a mass profile for the inner region of NGC 4636 using the
Chandra data. It agrees well with other profiles \citep{mus94, bri97, loe03} for $4<r<40$ kpc, while it is much steeper than the others for $r<4$ kpc.

\citet{mat98} also derived a total mass profile of NGC 4636 
  using {\it ASCA} GIS observations, 
  but  they did not present the information on the parameters of the mass profile.
So we read the mass profile in their Figure 3 and re-derive the mass profile in the functional form  
  following the approach of \citet{cot03} for M49.
\citet{cot03} considered that the total mass $M_{\rm tot}(r)$ interior to any radius is 
  the sum of dark matter mass $M_{\rm dm}(r)$ 
  (eq. \ref{eq-dmmass}) and stellar mass $M_{\rm s}(r)$ (eq. \ref{eq-starmass}).
If the dark matter density profile is the NFW halo model, $\rho_{\rm dm} (r)=K(r/r_s)^{-1}(1+r/r_s)^{-2}$, 
  the mass profile for dark matter halo is represented by

\begin{eqnarray}
M_{\rm dm}(r)   & = &\int_{0}^{r}{4{\pi}x^2\,\rho_{\rm dm}(x)}dx \nonumber \\
                & = & 4{\pi}Kr^{3}_{s}\left[\ln \left(1+{r\over r_s}\right)-{(r/r_s)\over 1+(r/r_s)}\right]\ ,
\label{eq-dmmass}
\end{eqnarray}
where $K$ is the dark matter density normalization and $r_s$ is a scale length.

Fixing $M_{\rm s}(r)$ with mass-to-light ratio $\Upsilon_0=9.0~M_\odot L^{-1}_{R,\odot}$,
  we fit the mass profile given by \citet{mat98}
  with the total mass profile that is the sum of dark matter mass $M_{\rm dm}(r)$ 
  (eq. \ref{eq-dmmass}) and stellar mass $M_{\rm s}(r)$ (eq. \ref{eq-starmass}).
Thus we determine $K=4.01\times10^{5}~M_\odot$kpc$^{-3}$ and $r_s=147$ kpc. 

Then the final mass models of NGC 4636 are
\begin{eqnarray}
M_{\rm tot}(r) & = & M_{\rm s}(r) + M_{\rm dm}(r), \nonumber \\
M_{\rm s}(r)   & = & 4.87\times 10^{11} M_\odot 
    \left[{(r/2.61\,{\rm kpc})^{1/2} \over 1+(r/2.61\,{\rm kpc})^{1/2}}\right]^{4.33} \nonumber \\
 & &   \times \left[{5.33+(r/2.61\,{\rm kpc})^{1/2} \over 1+(r/2.61\,{\rm kpc})^{1/2}}\right], \nonumber \\
M_{\rm dm}(r) & = & 1.61\times 10^{13} M_\odot \nonumber \\
 & &   \times \left[\ln \left(1+{r\over 147\,{\rm kpc}}\right)-{(r/147\,{\rm kpc})\over (1+r/147\,
    {\rm kpc})}\right].\ 
\label{eq-totalmass}
\end{eqnarray}

In Figure \ref{fig-mass}, we plot the stellar mass profile (dotted line), 
  the dark matter halo (long-dashed line), 
  and the total mass profile (dot-dashed line) derived from this method. 
  
While four mass profiles given by \citet{mus94, bri97, loe03, joh09} show monotonic increase
  with increasing radius, only the mass profile given by \citet{mat98} shows a flattening at
  $r\approx 10-20$ kpc, and then keeps increasing thereafter out to $r\approx400$ kpc.
From this comparison we conclude that the mass profile for $r>10$ kpc given by \citet{mat98} may be in error.
Since X-ray mass profiles that account for the dark matter halo are similar to or
  larger than the stellar mass profile at $r>2$ kpc,
  we consider that the X-ray mass profiles are good enough to determine
  the velocity anisotropy of the GC system for the following analysis.

\subsubsection{Determination of the Velocity Anisotropy}

The velocity anisotropy of GCs is determined as follows.
First, assuming the velocity anisotropy [$\beta_{\rm cl}(r)$] in prior,
  we derive the theoretical projected VDP [$\sigma_p(R)$] and theoretical
  projected aperture VDP [$\sigma_{ap}(\le R)$] using the Jeans equation.
The projected aperture VDP is the velocity dispersion of all objects 
  interior to a given projected radial distance $R$.
To obtain those theoretical VDPs, 
  we use the GC number density profile [$n_{\rm cl}(r)$]
  of the combined, blue, and red GCs and the mass profile [$M_{\rm tot} (r)$] 
  derived in the previous section.
Second, from the comparison of these calculated VDPs with measured VDPs,
  we determine the velocity anisotropy of GCs.

We begin by deriving the theoretical projected VDPs.
The  spherical Jeans equation (eq. \ref{eq-jeans}) can
be solved for the radial component of velocity dispersion, $\sigma_r(r)$:
\begin{eqnarray}
\sigma_r^2(r) 
  &=& {1\over{n_{\rm cl}(r)}}\, \exp\left( -\int{{2\beta_{\rm cl}}\over{r}}\,dr \right)\, \nonumber \\
  & & \times \left[ \int_r^{\infty} n_{\rm cl}\,{{GM_{\rm tot}}\over{x^2}}\, 
         \exp\left( \int{{2\beta_{\rm cl}}\over{x}}\,dx \right)\, dx \right]\ .
\label{eq-sigr}
\end{eqnarray}

Then the projected VDP, $\sigma_p(R)$ can be derived by
\begin{equation}
\sigma_p^2(R) = {2\over{N_{\rm cl}(R)}}\, \int_R^{\infty} n_{\rm
cl} \sigma_r^2(r) \left(1 - \beta_{\rm cl}\,{{R^2}\over{r^2}}
\right)\, {{r\,dr}\over{\sqrt{r^2 - R^2}}}, 
\label{eq-sigp}
\end{equation}.

The projected aperture VDP, $\sigma_{\rm ap}(\le R)$, 
  can be computed by
\begin{eqnarray}
\sigma_{\rm ap}^2(\le R) 
  &=& \left[ \int_{R_{\rm min}}^R N_{\rm cl}(R^\prime) \sigma_p^2(R^\prime)\,
         R^\prime\,dR^\prime \right]\, \nonumber \\
  & & \times \left[\int_{R_{\rm min}}^R N_{\rm cl}(R^\prime)\,R^\prime\,dR^\prime
         \right]^{-1}\ ,
\label{eq-apsig}
\end{eqnarray}
\noindent where $R_{\rm min}$ is the projected galactocentric distance of the
  innermost data point in the GC sample ($R_{\rm min}=1.63$ kpc in this study).

\begin{figure}
\plotone{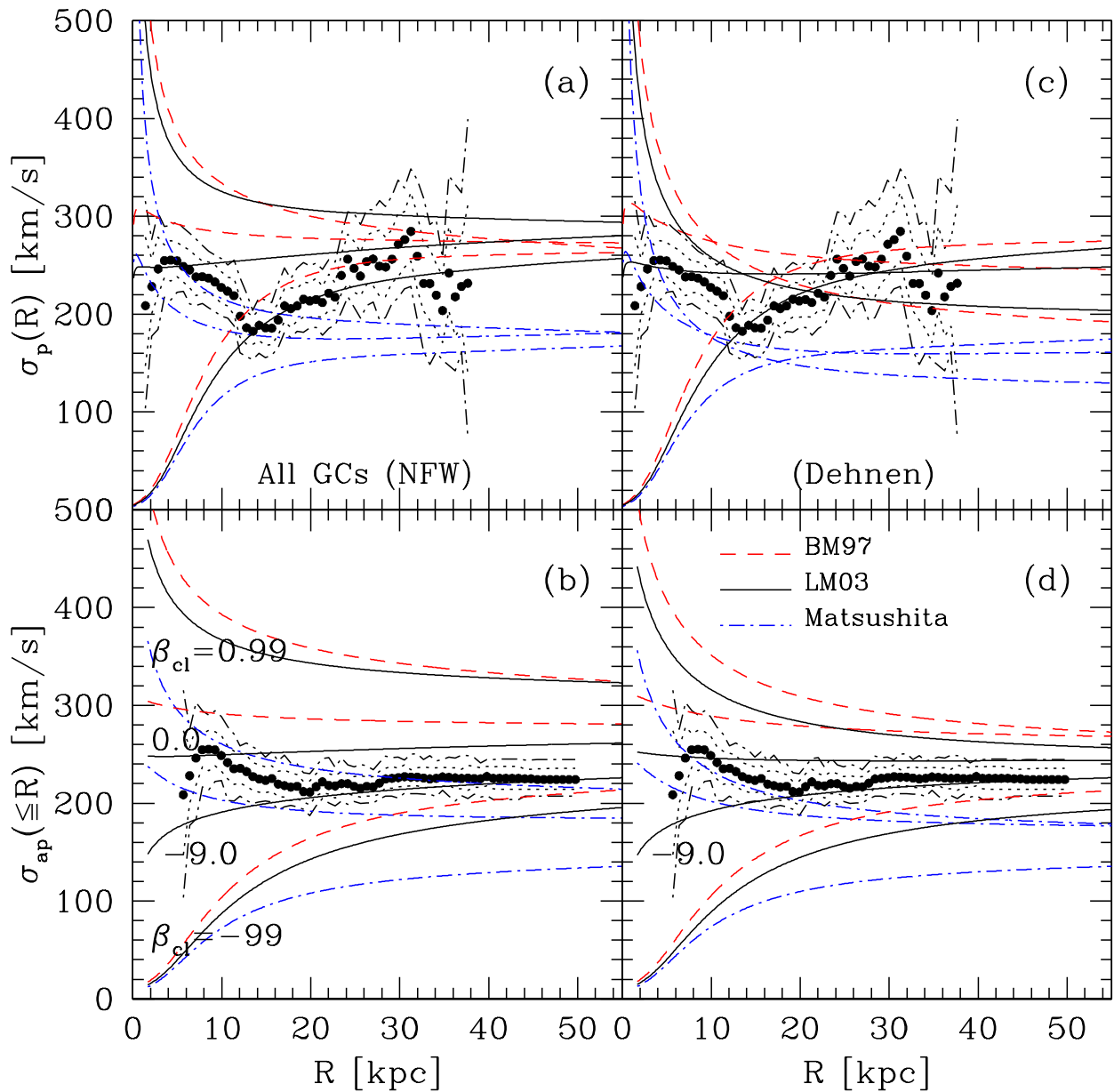} 
\caption{VDPs (a, c) and the aperture VDPs (b, d) for all the GCs.
Filled circles in (a, c) represent the measured VDP shown in Fig. \ref{fig-disp},
  and those in (b, d) denote the measured aperture VDP.
Associated, dotted and dot-dashed lines represent 68\% and
  95\% confidence intervals on the calculation of velocity dispersion, respectively.
Three smoothly curved lines, from a radially biased
  velocity anisotropy to the tangentially biased velocity anisotropy
  (from top to bottom, $\beta_{\rm cl}$= 0.99, 0, and $-$99),
  represent the VDPs calculated using
  the GC density profile of NFW (a,b) and of Dehnen (c,d). 
The solid lines represent the VDPs calculated using the total mass profile derived from \citet{loe03},
and the dashed and the dot-dashed lines represent the mass profile by \citet{bri97} and \citet{mat98}, respectively. The solid lines labeled with $\beta_{\rm cl} =-9.0$ in (b) and (d) represent approximate fits for $R>20$ kpc.
\label{fig-isoagc}}
\end{figure}

\begin{figure}
\plotone{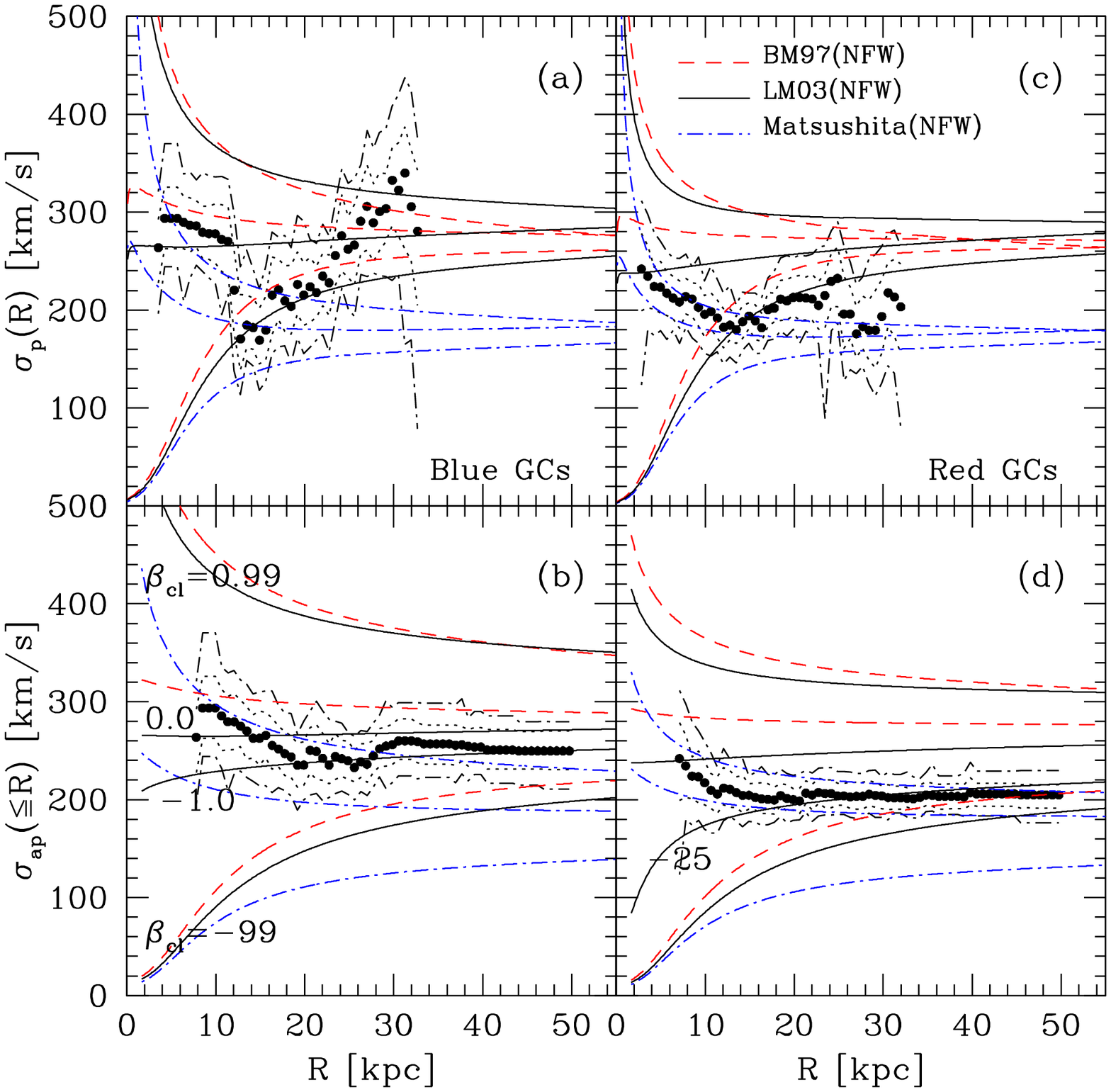} 
\caption{
VDPs (a, c) and the aperture VDPs (b, d) for the blue and red GCs.
The symbols are the same as in Fig. \ref{fig-isoagc}, but 
the dashed, the solid, and the dot-dashed lines represent the VDPs calculated using the mass profile 
derived from \citet{bri97}, \citet{loe03}, and \citet{mat98} with the GC density profile of NFW, respectively. The solid lines labeled with $\beta_{\rm cl} =-1.0$ and --25 in (b) and (d), respectively, represent approximate fits for $R>20$ kpc.
\label{fig-isobgc}}
\end{figure}

We present the measured VDP in comparison with the
  VDPs calculated by assuming several velocity anisotropies
  in Figures \ref{fig-isoagc} and \ref{fig-isobgc}. 
The left and right panels in Figure \ref{fig-isoagc} show the VDPs calculated using
  the NFW profile and the Dehnen for the GC number density, respectively. 
The upper panels show the projected VDPs, and the lower panels
show the projected aperture VDPs. The measured dispersion data
taken from Figure \ref{fig-disp} are shown by filled circles along
with their confidence intervals. The projected aperture VDPs in
  the lower panels are plotted in the similar fashion to the case
  of the upper panels.

The mean velocity dispersion for the outer part shows a difference  by
a factor of $\sim 1.5$ 
between the VDP profiles based on \citet{mat98} and those on \citet{bri97} and \citet{loe03}.
This is because the mass in the outer part derived from \citet{bri97} and \citet{loe03} 
  is about twice larger than that from \citet{mat98}. 
As a result, the system of all the GCs
has a tangentially biased velocity ellipsoid ($\beta_{\rm cl}<0$) 
  for the mass profiles by \citet{bri97} and \citet{loe03},
but has a radially biased velocity ellipsoid ($\beta_{\rm cl}>0$) for the \citet{mat98} mass profile, 
  in Figure \ref{fig-isoagc} (a,b) based on the NFW profile for the GC number density.
A similar result can be found in Figure \ref{fig-isoagc} (c,d) based on
  the Dehnen profile for the GC number density.
We tried an eye-ball fit to the data for the outer region of NGC 4636, 
finding that the VDP profile with $\beta_{\rm cl} =-9.0$ (derived for the \citet{loe03} mass profile)
fits approximately the data for $R>20$ kpc (shown by the solid lines labeled with $\beta_{\rm cl} =-9.0$ in (b) and (d)).
 
In Figure \ref{fig-isobgc}, 
  we show a similar analysis for the blue (left panels) and red GCs (right panels) based on the NFW profile for the GC number density. 
These figures show similar trends to those for all the GCs.
They have a tangentially biased velocity ellipsoid ($\beta_{\rm cl}<0$) 
  for the \citet{bri97} and \citet{loe03} mass profiles,
while they  have a radially biased velocity ellipsoid ($\beta_{\rm cl}>0$) for the \citet{mat98} mass profile. 
When we compare the velocity anisotropies between the blue and red GCs by fixing the mass profile, 
we find the velocity anisotropies of the red GCs by the \citet{bri97} and \citet{loe03} mass profiles 
  are slightly more tangential than those of the blue GCs in Figure \ref{fig-isobgc}.
These trends are similar to the velocity anisotropies based on the Dehnen profile for the GC number density, too, although they were not plotted in Figure \ref{fig-isobgc}. 
We tried an eye-ball fit to the data for $R>20$ kpc, 
finding that the VDP profiles with $\beta_{\rm cl} =-1.0$ and --25 (derived for the \citet{loe03} mass profile), fit approximately the data for the blue GCs and red GCs, respectively (shown by the solid lines labeled with $\beta_{\rm cl} =-1.0$ and --25, respectively, in (b) and (d)).
It is noted that the velocity dispersion for the blue GCs increases as $R$ decreases so that the 
orbit of the blue GCs becomes radial in the inner region at $R<12$ kpc.  

Since we concluded that the mass profile given by \citet{mat98} may be in error before, we adopt finally the results based on the mass profile given by \citet{loe03} (that is also similar to that given by
\citet{bri97}).
In summary, it is found that
 the orbit of the GC system in NGC 4636 is tangential. 
 The orbits of both the red GCs and blue GCs are tangential, and the orbits of 
  the red GCs is slightly more tangential than that of the blue GCs.
   
\section{Discussion}\label{discuss}

\subsection{GC Kinematics and the Global Properties of gEs}\label{global}

In this section, we combine the kinematic properties of the GC systems in gEs by including 
  the results for the NGC 4636 GC system in this study,
  those for the NGC 1407 GC system (172 GCs) given in \citet{rom09},
  and those for the GC systems in other gEs in \citet{hwa08}.
  \citet{hwa08} analyzed the velocity data of 276 GCs in M87 \citep{cot01}, 
  263 GCs in M49 \citep{cot03}, 435 GCs in NGC 1399 \citep{ric04}, 
  341 GCs in NGC 5128 \citep{woo07}, 
  121 GCs in M60 \citep{lee08a}, and 172 GCs in NGC 4636 \citep{sch06},
  using the similar method adopted in this study (see also \citealt{rom09}).
We derived the kinematic parameters of the NGC 1407 GCs from the data in the catalog 
given by \citet{rom09} using the same analysis as used for other gEs in this study, 
and listed the results in Table \ref{tab-n1407kin}.
The information on the host galaxies is listed in Table \ref{tab-gesample},
and the global kinematics of GCs in gEs are summarized in Table \ref{tab-gekin}. 

The velocity dispersion of the GCs in NGC 4636 is similar to 
  that of the M60 GCs ($\sigma_p=234^{+ 13}_{-14}$ km s$^{-1}$),
  although NGC 4636 is about one magnitude fainter than M60.
This indicates that the mass to luminosity ratio is larger in NGC 4636 than in M60.
The velocity dispersion of the GCs in NGC 4636 is much smaller than those for the three
  brightest gEs (M87, M49 and NGC 1399).
However, the velocity dispersion of the GCs in NGC 4636 is about twice larger than
  than that of NGC 5128 GCs ($\sigma_p=129^{+ 5}_{-7}$ km s$^{-1}$), 
  although NGC 4636 is slightly fainter than NGC 5128.

The rotation-corrected velocity dispersion,
  $\sigma_{p,r}$ of the blue GCs in NGC 4636 is larger than that of
  the red GCs. This trend is similarly seen in the three X-ray brightest gEs.
However it is opposite in the case of M60, and the rotation-corrected velocity dispersions
of both blue and red GCs are similar in NGC 5128 and NGC 1407.
The rotation-corrected velocity dispersion ranges from 129 km s$^{-1}$ (NGC 5128) to
  399 km s$^{-1}$ (M87) for all the GCs, and similarly for the blue GCs and red GCs. 
The rotation amplitude ranges from 30 km s$^{-1}$ (NGC 5128) to
  172 km s$^{-1}$ (M87) for all the GCs, and similarly for the blue GCs and red GCs.
The ratio of the rotation amplitude to the rotation-corrected velocity dispersion ranges
  from 0.10 (NGC 1399) to 0.65 (M60). 
M87 and M60 show much stronger rotation with $\Omega R /\sigma_{p,r}^{AGC}>0.4$ than the others.

We investigate any dependence of the kinematic properties of the GC systems in gEs 
on the global properties of their host galaxies.
  We calculated the Spearman's correlation coefficient ($r_S$) and its significance ($\sigma(r_S )$)
  to check any correlation  between parameters, and used the bisector method \citep{iso90} 
  to do linear fits for correlated pairs of parameters.

\begin{figure}
\epsscale{.9}
\plotone{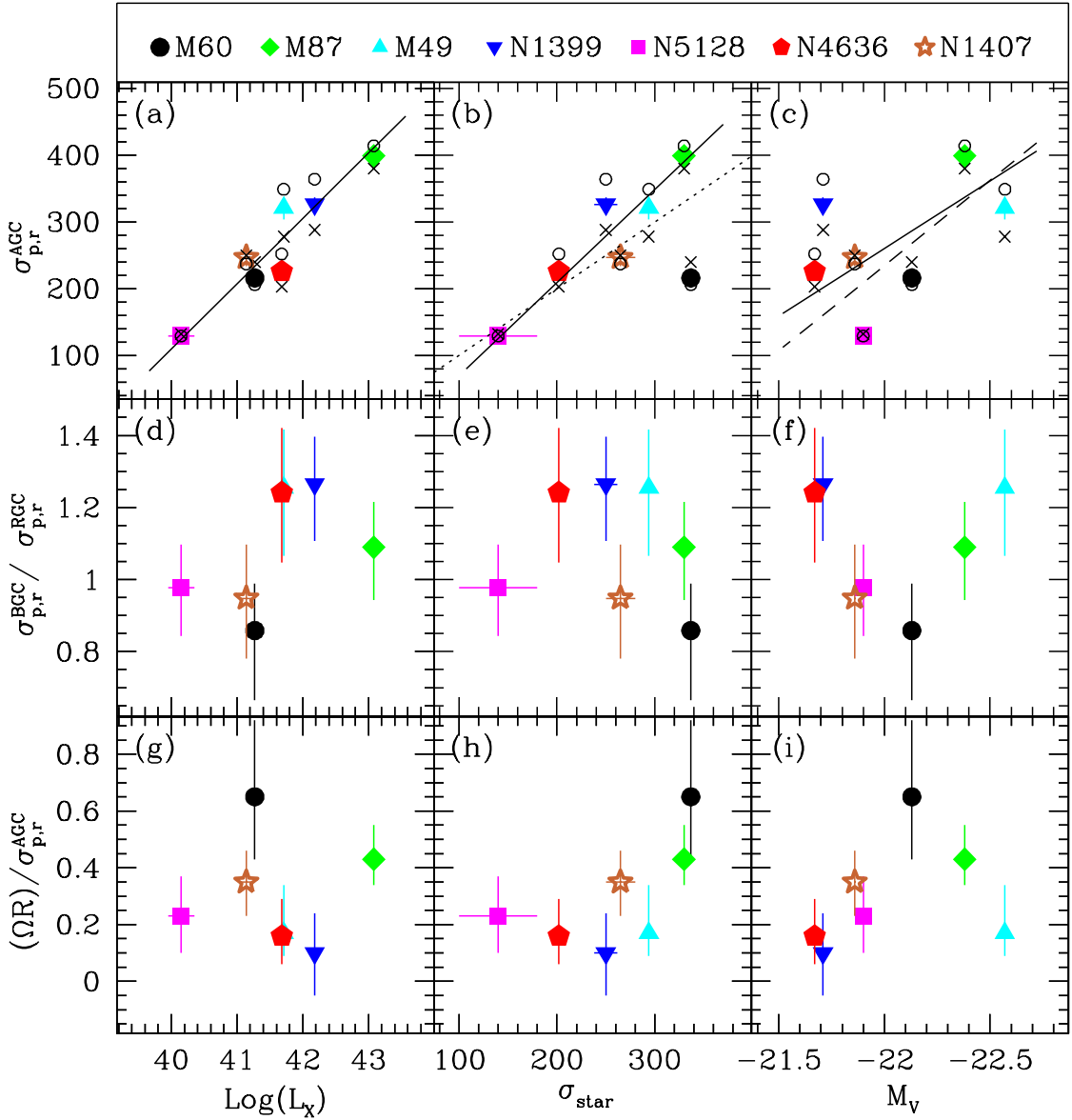}
\caption{Kinematic parameters (the velocity dispersion for all the GCs,  $\sigma_{p,r}^{AGC}$,
  the ratio of the velocity dispersion between the blue GCs and red GCs, $\sigma_{p,r}^{BGC}/\sigma_{p,r}^{RGC}$,
  the ratio of the rotational velocity to the velocity dispersion for all the GCs,
 $\Omega R /\sigma_{p,r}^{AGC}$)
  vs. global parameter 
  (X-ray luminosity, $\log L_X$, stellar velocity dispersion, $\sigma_{star}$, 
   and total $V$-band magnitude, $M_V$) for  the gEs: 
  M60 (circles), M87 (diamonds), M49 (triangles), NGC 1399 (reversed triangles), 
  NGC 5128 (squares),  NGC 4636 (pentagons), and NGC 1407 (stars).
  Open circles and crosses in (a), (b), and (c) represent the blue and red GCs
  in each gE, respectively.
The solid lines represent the bisector linear fits. 
The dotted line in (b) represents one-to-one relation.
The dashed line in (c) represents the fit without NGC 1399, which is listed in Table 5.
 \label{fig-kinmv}}
\end{figure}

Figure \ref{fig-kinmv} displays three kinematic parameters 
  (the rotation-corrected velocity dispersion for all the GCs, $\sigma_{p,r}^{AGC}$  (as well as the blue and red GCs),
  the ratio of the rotation-corrected velocity dispersion
  between the blue GCs and red GCs, $\sigma_{p,r}^{BGC}/\sigma_{p,r}^{RGC}$,
  and the ratio of the rotational velocity to the rotation-corrected velocity dispersion for all the GCs,
  $\Omega R /\sigma_{p,r}^{AGC}$) versus three global parameters 
  (X-ray luminosity, log $L_X$, stellar velocity dispersion, $\sigma_{star}$, and total $V$-band magnitude, $M_V$)
  for the gEs, where the rotation-corrected velocity dispersion means the velocity dispersion about the best fit rotation curve.
Here the stellar velocity dispersion represents the mean value of the velocity stellar dispersion at $\sim R_{\rm eff}/4$ for given galaxy: NGC 1399 \citep{sag00}, NGC 5128 \citep{wil86}, NGC 1407 \citep{spo08} and other galaxies \citep{ben94} .

The rotation-corrected velocity dispersion for all the GCs shows a
  strong correlation with the X-ray luminosity and stellar velocity dispersion, and a weaker correlation with the total $V$-band magnitude, as seen in Figure \ref{fig-kinmv}(a), (b) and (c).
This shows that the rotation-corrected velocity dispersion for the GCs is an excellent indicator
  for the mass or luminosity of their host galaxies.
It is noted that the stellar velocity dispersion of M60 is as large as that of M87 
  (see Fig. \ref{fig-kinmv}(b)), being much larger than that expected from its luminosity.
This may be related with the presence of companion SBc galaxy NGC 4647 located at $2\arcmin'.5$ 
from M60,
but the cause of this large stellar velocity dispersion of M60 is not known.
M60 seems to be different from the other gEs in stellar velocity dispersion. 
Therefore we excluded any parameters for M60 related with the stellar velocity dispersion 
  for the following correlation analysis.

Linear fits to the data yield:
$\sigma_{p,r}^{AGC} = 97.66 \log L_X - 3796$ with rms = 28 km s$^{-1}$,
$\sigma_{p,r}^{AGC} = 1.394 \sigma_{star} - 69.51$ with rms = 31 km s$^{-1}$, and
$\sigma_{p,r}^{AGC} = -202.8 M_V - 4201$ with rms = 76 km s$^{-1}$.
In the case of $\sigma_{p,r}$ and $\log L_X$, 
the blue GCs show a stronger correlation than the red GCs:
$\sigma_{p,r}^{BGC} = 110.28 \log L_X - 4309$ with rms = 32 km s$^{-1}$ and the Spearman's rank correlation coefficient $r_S=0.96$ for the blue GCs, 
and $\sigma_{p,r}^{RGC} = 84.04 \log L_X - 3243$ with rms = 27 km s$^{-1}$ and $r_S=0.86$ for the red GCs. It is noted that the slope for the blue GCs (110.28) is steeper than that for the red GCs (84.04).
On the other hand, it is opposite in the case of $\sigma_{p,r}$ and $\sigma_{star}$. 
The red GCs show a stronger correlation than the blue GCs:
$\sigma_{p,r}^{RGC} = 1.237 \sigma_{star} - 50.31 $ with rms = 24 km s$^{-1}$ and 
$r_S=0.0.83$ for the red GCs, and $\sigma_{p,r}^{BGC} = 1.538 \sigma_{star} - 88.77$ with rms = 46 km s$^{-1}$ and $r_S=0.71$ for the blue GCs. It is noted that the slope for the blue GCs (1.538) is steeper than that for the red GCs (1.237).

The slope for the relation between $\sigma_{p,r}^{AGC}$ and $\sigma_{star}$
  (solid line) is larger than one (dotted line). 
This indicates that the more massive gEs are, the more massive dark matter halo they have.
The relation between $\sigma_{p,r}^{AGC}$ and $M_V$ shows a larger scatter than
 that between  $\sigma_{p,r}^{AGC}$ and X-ray luminosity. 
However, NGC 1399 has an absolute magnitude much fainter than expected from its X-ray luminosity. 
If NGC 1399 is removed, the linear fit yields $\sigma_{p,r}^{AGC} = -255.6 M_V - 5389$ 
with a smaller scatter, rms = 65 km s$^{-1}$ (dashed line in Fig. \ref{fig-kinmv}(c)). 
On the other hand, little correlation is seen between other pairs of parameters.

\begin{figure}
\epsscale{.9}
\plotone{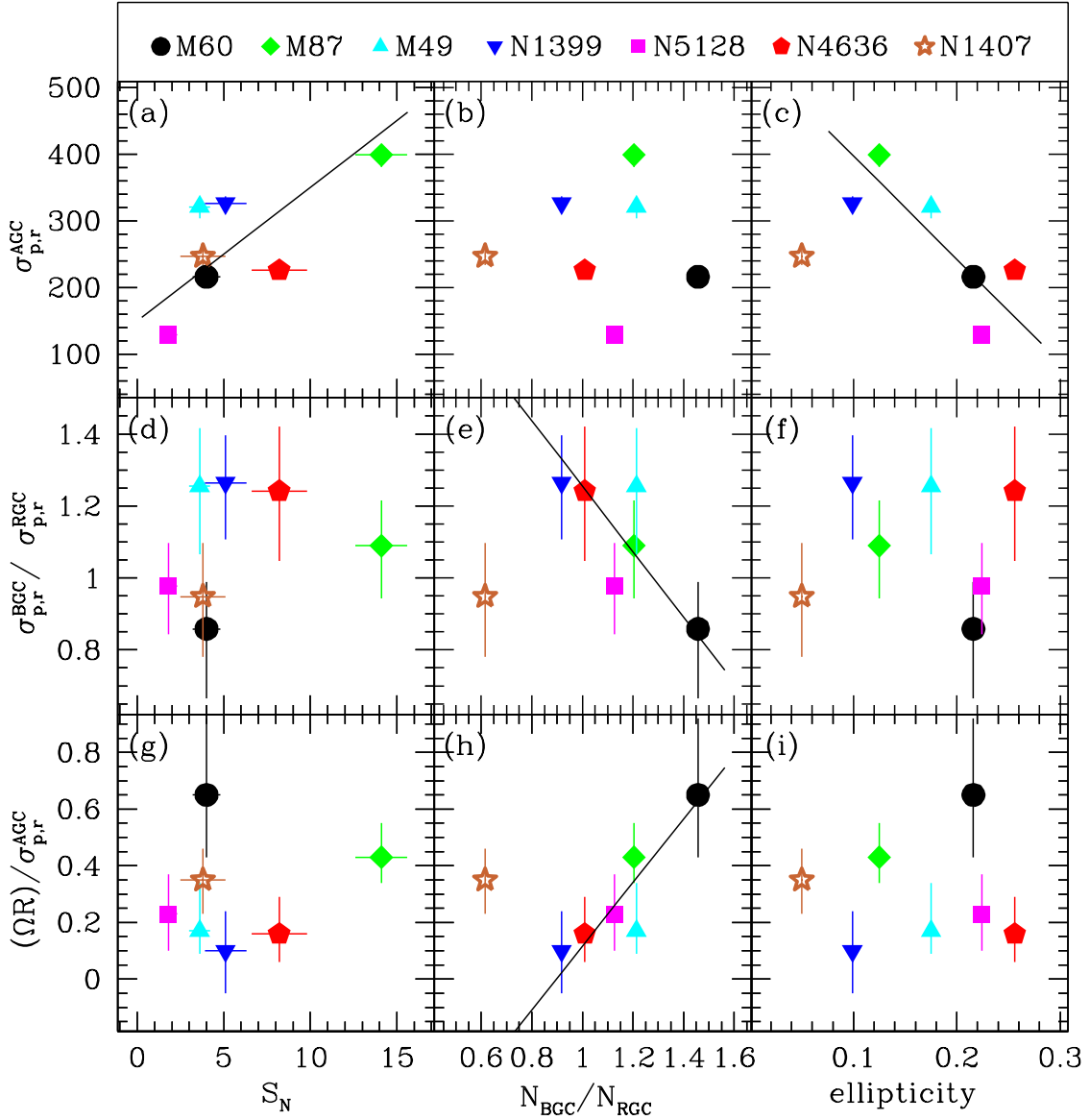} 
\caption{
Kinematic parameters (the velocity dispersion for all the GCs, $\sigma_{p,r}^{AGC}$ ,
  the ratio of the velocity dispersion between the blue GCs and red GCs, $\sigma_{p,r}^{BGC}/\sigma_{p,r}^{RGC}$,
  the ratio of the rotational velocity to the velocity dispersion for all the GCs)
  vs. global parameters (specific frequency, $S_N$, 
the number ratio of the blue GCs and the red GCs,   $N_{BGC}/N_{RGC}$, 
and ellipticity,$\epsilon$) for the gEs.
The symbols are the same as in Fig. \ref{fig-kinmv}.
The solid lines represent the bisector linear fits. 
\label{fig-kinsn}}
\end{figure}

In Figure \ref{fig-kinsn}, we display the same three kinematic parameters 
  versus three other global parameters
  (specific frequency, $S_N$, the number ratio of the blue GCs and the red GCs, $N_{BGC}/N_{RGC}$, 
  and ellipticity, $\epsilon$) for the gEs.
We find strong correlations for some pairs of parameters:
(a) $\sigma_{p,r}^{AGC}$ and $S_N$, 
(b) $\sigma_{p,r}^{AGC}$ and $\epsilon$, 
(c) $\sigma_{p,r}^{BGC}/\sigma_{p,r}^{RGC}$ and $N_{BGC}/N_{RGC}$, and
(d) $\Omega R /\sigma_{p,r}^{AGC}$ and $N_{BGC}/N_{RGC}$.
Linear fits to the data yield:
$\sigma_{p,r}^{AGC} = 20.03 S_N + 150$ with rms = 63 km s$^{-1}$ (panel a), 
$\sigma_{p,r}^{AGC} = -1548 \epsilon + 552$ with rms = 57 km s$^{-1}$ (panel c),
$\sigma_{p,r}^{BGC}/\sigma_{p,r}^{RGC} = -0.908 N_{BGC} / N_{RGC} + 2.16$ 
with rms = 0.11 (panel e), and
$\Omega R /\sigma_{p,r}^{AGC} = 1.120 N_{BGC} / N_{RGC} - 1.00$ with rms = 0.09 (panel h).
NGC 1407 shows much different relations including $N_{BGC} / N_{RGC}$ from those of other gEs so that
we did not use NGC 1407 for linear fitting for (c), (e) , and (h). 
Thus the rotation-corrected velocity dispersion for all the GCs increases as $S_N$ increases,
  but this velocity dispersion decreases as ellipticity increases.
$\sigma_{p,r}^{BGC}/\sigma_{p,r}^{RGC}$ decreases as  $ N_{BGC} / N_{RGC}$ increases, 
while $({\Omega R}) / \sigma_{p,r}^{AGC}$ increases as  $ N_{BGC} / N_{RGC}$ increases.

\begin{figure}
\plotone{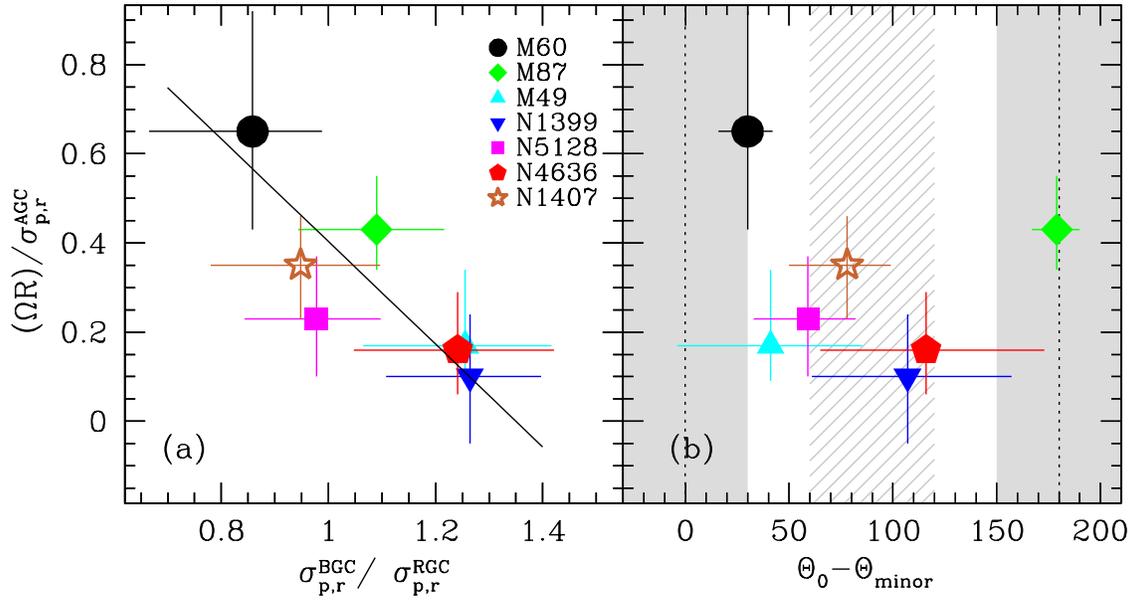} 
\caption{The ratio of the rotational velocity to the velocity dispersion for all the GCs,
$\Omega R /\sigma_{p,r}^{AGC}$, vs. 
the ratio of the velocity dispersion between the blue GCs and red GCs,  $\sigma_{p,r}^{BGC}/\sigma_{p,r}^{RGC}$ (a), 
and the difference between the GC rotation angle and
the position angle of the minor axis of the host galaxies, $\Theta_0 - \Theta_{minor}$ (b) 
  for the gEs. 
The symbols are the same as in Fig. \ref{fig-kinmv}.
The solid lines represent the bisector linear fits. 
\label{fig-disprot}}
\end{figure}

Figure \ref{fig-disprot}(a) displays directly the relation between the velocity dispersion and 
  rotational amplitudes of the GCs: 
The ratio of the velocity dispersion between the blue GCs and red GCs  ($\sigma_{p,r}^{BGC}/\sigma_{p,r}^{RGC}$) vs.
  the ratio of the rotational velocity to the velocity dispersion for all the GCs
($\Omega R /\sigma_{p,r}^{AGC}$). 
 It is found that $\sigma_{p,r}^{BGC}/\sigma_{p,r}^{RGC}$ 
  has a strong correlation with $\Omega R /\sigma_{p,r}^{AGC}$:
$\Omega R /\sigma_{p,r}^{AGC} = -1.153 \sigma_{p,r}^{BGC}/\sigma_{p,r}^{RGC} + 1.556$ with rms=0.108.
That is, the weaker the rotation is, the larger the velocity dispersion ratio 
 between the blue GCs and red GCs is.

Figure \ref{fig-disprot}(b) displays directly the relation between the rotational amplitudes and 
  rotational axis of the GCs: 
The ratio of the rotational velocity to the velocity dispersion for all the GCs
($\Omega R /\sigma_{p,r}^{AGC}$) vs. 
 the difference between the GC rotation angle and
the position angle of the minor axis of their host galaxies ($\Theta_0 - \Theta_{minor}$).
Interestingly it is seen that  the GC rotation axis for the two gEs with the strongest rotation (M60 and M87) 
is approximately aligned with the minor axis of their host galaxies (the dotted lines in the shaded regions).
This shows that the GC systems in gEs rotate around the minor axis of their host galaxies, once they
have some strong rotation.

In Table \ref{tab-relation}, we summarized the results of the linear fits ($Y=aX+b$) given in 
  Figures \ref{fig-kinmv}, \ref{fig-kinsn}, and \ref{fig-disprot}.
We listed the Spearman's rank correlation coefficient ($r_S$) and its significance ($\sigma(r_S )$) in the 6th and 7th columns , respectively.   
These correlations found in this study provide strong constraints on modeling the GC systems in gEs (to be discussed in \S \ref{models}). 
Moreover, it is needed, in the future studies, to investigate which relation is the most fundamental one, if any,  
  among several correlations between kinematic properties of the GC systems in gEs
  and the global parameters of their host galaxies found in this study.

\begin{figure}
\plotone{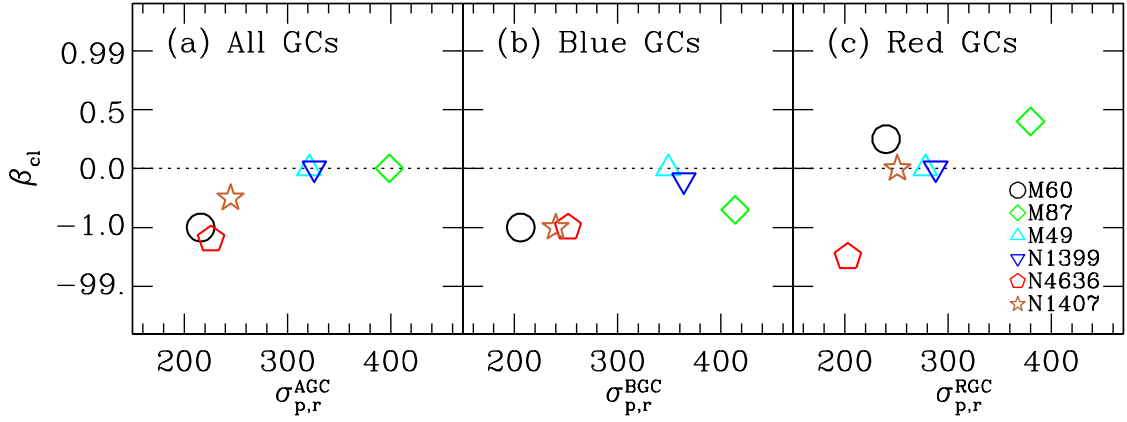} 
\caption{
The velocity anisotropy ($\beta_{cl}$) vs. the velocity dispersion of all the GCs (a), blue GCs (b), and red GCs (c) in gEs.
The velocity anisotropies only indicate representative positions
  according to the decision of each velocity ellipsoid.
\label{fig-dispiso}}
\end{figure}

Figure \ref{fig-dispiso} represents the velocity anisotropy
  versus the velocity dispersions for the GCs in 6 gEs (M60, M87, M49, NGC 1399, NGC 1407, and NGC 4636), determined in this study, \citet{hwa08} and \citet{rom09}.  
We consider $\beta_{\rm cl} = 0.5$, 0, --1, and --99, as the radial, isotropic, tangential, and strongly tangential velocity anisotropy, respectively.
In the case of NGC 4636, we used the result derived for 
\citet{loe03} mass profiles.
All the GC systems in three gEs (M49, M87 and NGC 1399) show isotropic orbits, while
those in two gEs (M60 and NGC 4636) show tangential orbits. It it noted also that
more massive gEs (with larger GC velocity dispersion) have isotropic orbits, while
less massive gEs have tangential orbits.
The blue and red GCs in three gEs (M49, NGC 1399 and NGC 4636) show similar orbits,
while those in two gEs (M60 and M87) show opposite orbits.

\subsection{Radial Variation of the GC Kinematics for gEs}\label{radial}

\begin{figure}
\epsscale{.7}
\plotone{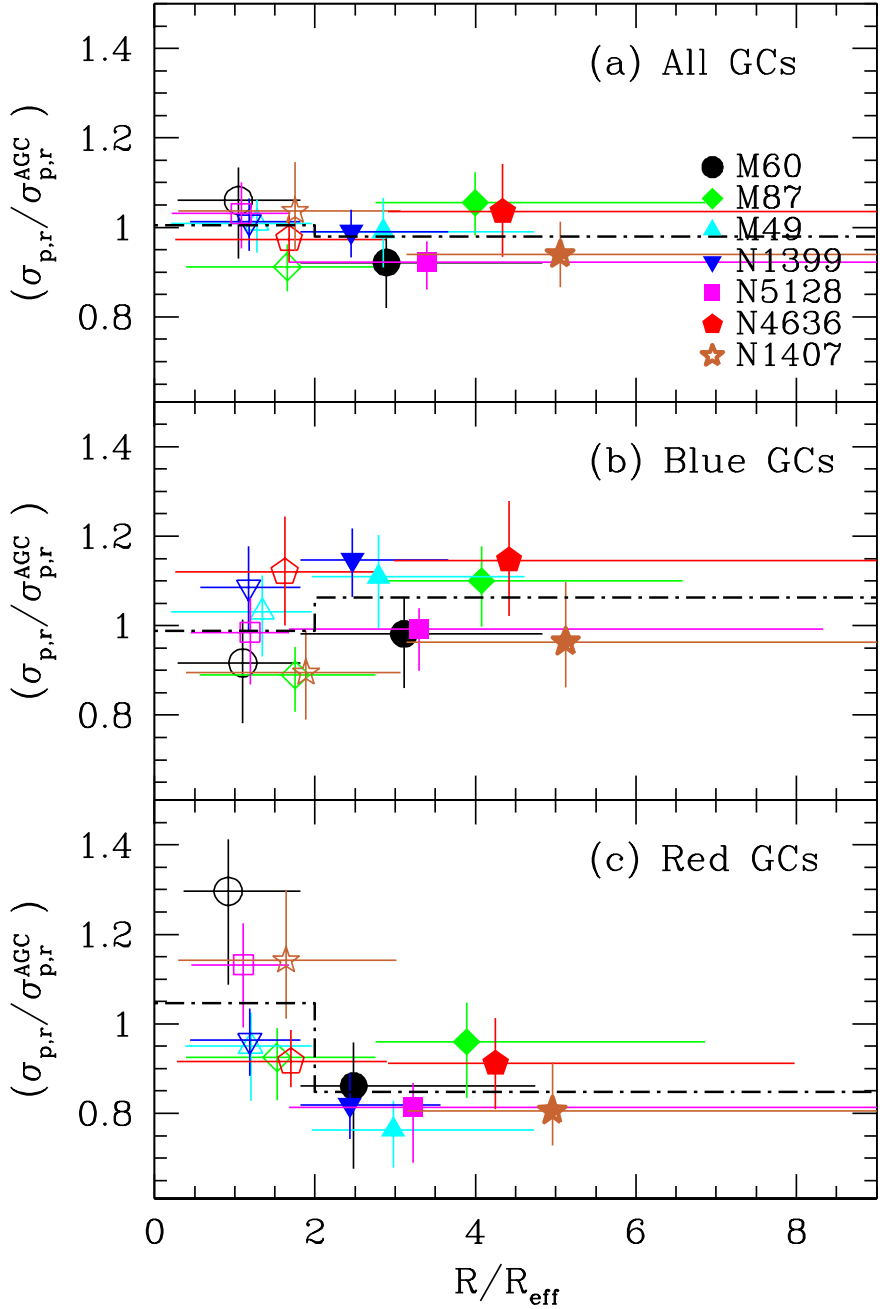} 
\caption{The rotation-corrected velocity dispersions, 
$\sigma_{p,r} / \sigma_{p,r}^{AGC}$,
vs. the projected galactocentric distances normalized to the effective radius, $R/R_{\rm eff}$,
 for all the GCs ({\it top}), blue GCs ({\it middle}), and red GCs ({\it bottom}) in the gEs.
Open symbols indicate the dispersions in the inner region of each gE, 
  while filled symbols those in the outer region.
The dot-dashed lines represent the average of the rotation-corrected velocity dispersions 
 for the inner region ($R/R_{\rm eff} <2$) and outer region ($R/R_{\rm eff} >2$) of each gE.
\label{fig-disprad}}
\end{figure}

To investigate the radial variation of the kinematics of the GC systems in the gEs,
  we plot the rotation-corrected velocity dispersions ($\sigma_{p,r} / \sigma_{p,r}^{AGC}$) 
  against the projected galactocentric distances in Figure \ref{fig-disprad}. 
The rotation-corrected velocity dispersion is normalized by that of all the GCs in each gE. 
The projected galactocentric distance is normalized with respect to the effective radius of each gE,
$R/R_{\rm eff}$.
The combined GCs do not show any significant change of
  the velocity dispersion over the whole region of a galaxy.
However, the mean velocity dispersion for the red GCs in the inner region ($R<2R_{\rm eff}$), 
 $1.05 \pm 0.15$, 
 is slightly larger than that in the outer region ($R>2R_{\rm eff}$),  $0.85 \pm 0.07$, 
  while that for the blue GCs in the inner region is little different from that in the outer region.
Note that the red GCs in M60, NGC 5128, and NGC 1407 have much larger $\sigma_{p,r} / \sigma_{p,r}^{AGC}$ in the inner
region  than in the outer region. Why the red GCs have larger velocity dispersion in the inner region than
in the outer region is an interesting question to solve.

\begin{figure}
\plotone{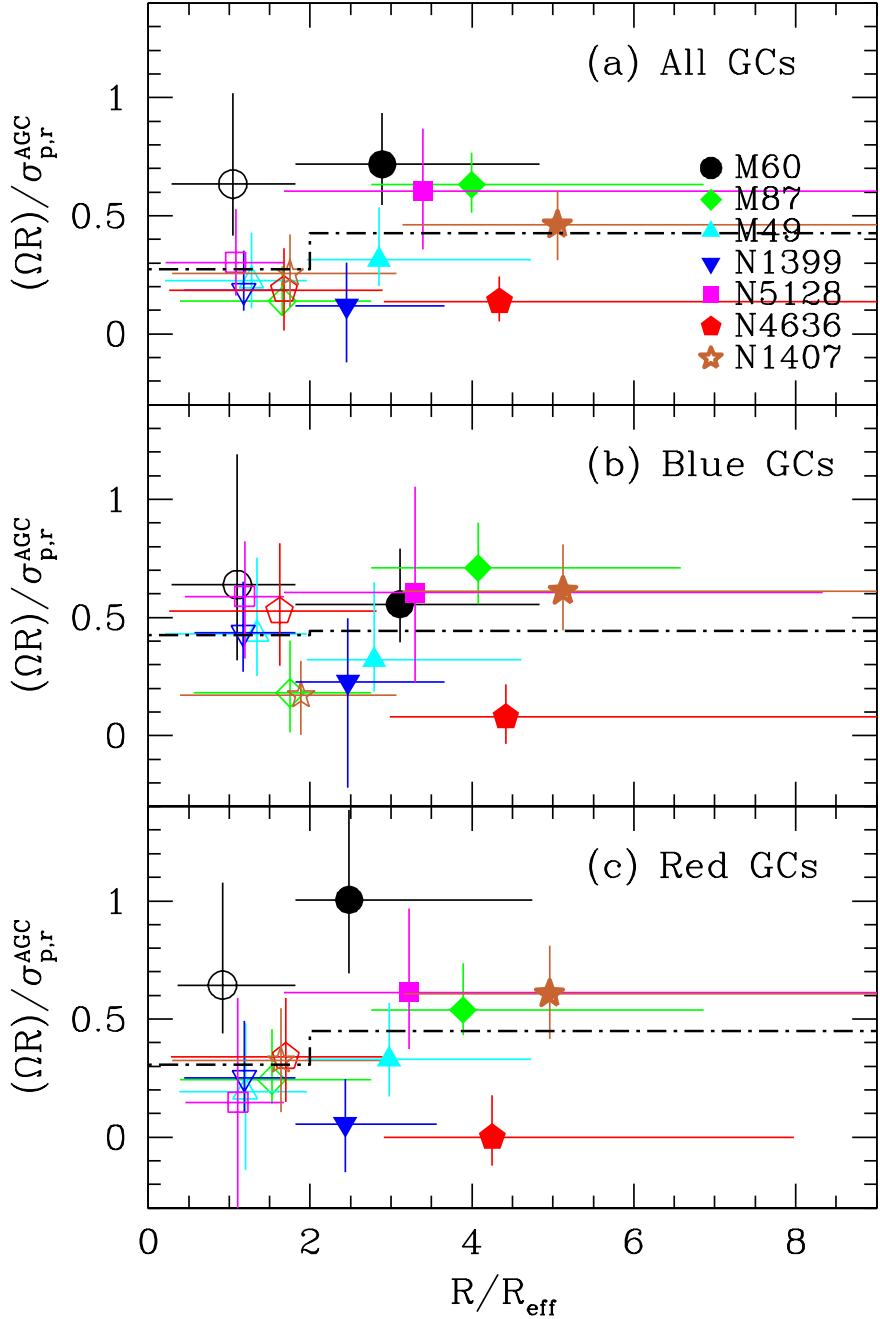}
\caption{
The ratio of the rotation amplitude to the velocity dispersion,  $\Omega R /\sigma_{p,r}^{AGC}$,
  vs. the projected galactocentric distances, 
$R/R_{\rm eff}$, 
  for all the GCs ({\it top}), the blue GCs ({\it middle}), and the red GCs ({\it bottom}) in gEs.
The symbols are the same as in Fig. \ref{fig-disprad}.
The dot-dashed lines represent the average of the rotation-corrected velocity dispersions 
 for the inner region and outer region of each gE.
  \label{fig-rotrad}}
\end{figure}

Figure \ref{fig-rotrad} displays 
 the ratio of the rotation amplitude to the rotation-corrected velocity
 dispersion ($\Omega R /\sigma_{p,r}^{AGC}$)  vs. the projected galactocentric distance 
normalized to the effective radius ($R/R_{\rm eff}$).
The combined GCs in M87, NGC 5128, and NGC 1407 show much larger $\Omega R /\sigma_{p,r}^{AGC}$ 
  in the outer region than in the inner region, 
  while those in other galaxies show little, if any, radial variation.
The blue GCs and red GCs in M87 and NGC 1407 show a similar trend, while only the red GCs in NGC 5128 show a similar trend.
However, in the case of NGC 4636, both the blue and red GCs show an opposite trend, 
a larger $\Omega R /\sigma_{p,r}^{AGC}$ 
  in the inner region than in the outer region. 
It is noted that the blue GCs and red GCs in NGC 4636 rotate in opposite direction to each other. 

\begin{figure}
\plotone{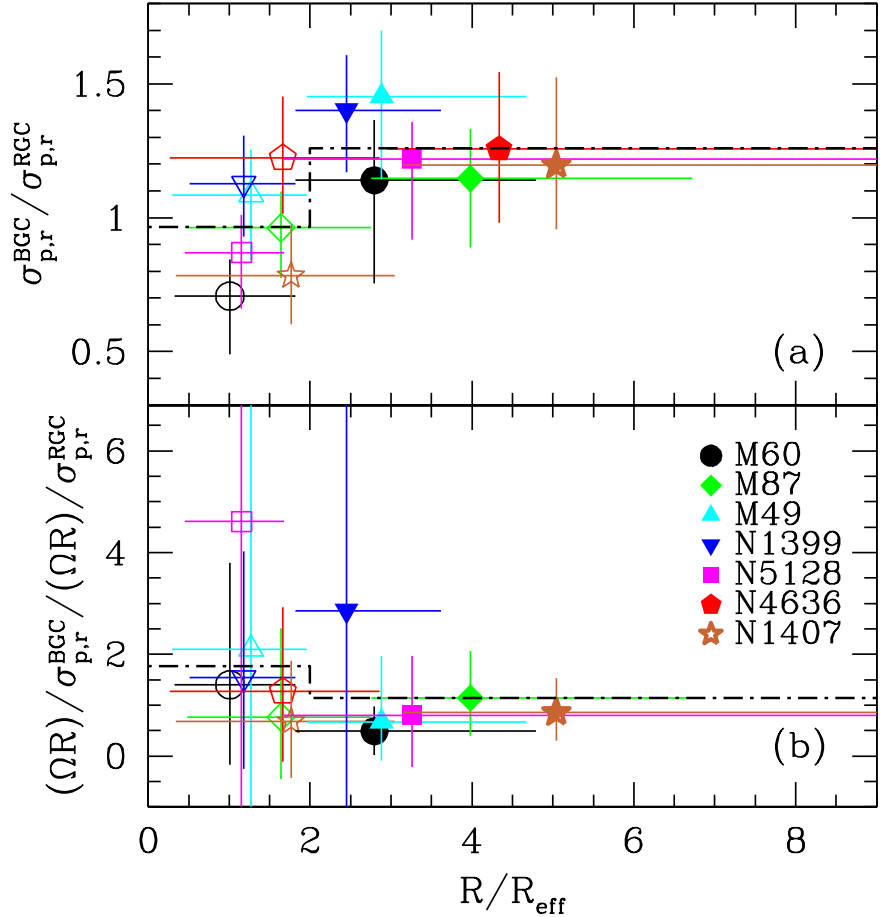} 
\caption{
The ratio of the rotation-corrected velocity dispersion between the blue GCs and the red GCs,
$\sigma_{p,r}^{BGC}/\sigma_{p,r}^{RGC}$  (a),  and
  the ratio of the rotation amplitude to the rotation-corrected velocity dispersion 
  between the blue GCs and the red GCs, 
$(\Omega R) /\sigma_{p,r}^{BGC} / (\Omega R) /\sigma_{p,r}^{RGC}$
 (b) vs. the projected galactocentric distance. 
The symbols are the same as in Fig. \ref{fig-disprad}.
The dot-dashed lines represent the average of the rotation-corrected velocity dispersions 
 for the inner region and outer region of each gE.
  \label{fig-disprotrad}}
\end{figure}

To show better the results seen in Figures \ref{fig-disprad} and \ref{fig-rotrad},
  we display in Figure \ref{fig-disprotrad} 
  the ratio of the rotation-corrected velocity dispersion between the blue GCs and red GCs
 ($\sigma_{p,r}^{BGC}/\sigma_{p,r}^{RGC}$)  and
  the ratio of the rotation amplitude to the rotation-corrected velocity dispersion 
  between the blue GCs and the red GCs ( $(\Omega R) /\sigma_{p,r}^{BGC} / (\Omega R) /\sigma_{p,r}^{RGC}$) 
as a function of projected galactocentric distance normalized to the effective radius
($R/R_{\rm eff}$).
It is seen that the mean value of $\sigma_{p,r}^{BGC}/\sigma_{p,r}^{RGC}$ 
  for the inner region ($R<2R_{\rm eff}$), is close to one, 0.97 $\pm$ 0.19,
  and that it is slightly smaller than that for the outer region ($R > 2R_{\rm eff}$), 
  1.26 $\pm$ 0.12.
It is also noted that the values of $(\Omega R) /\sigma_{p,r}^{BGC} / (\Omega R) /\sigma_{p,r}^{RGC}$ 
  for the inner region, 1.77 $\pm$ 1.34 are marginally larger than those for the outer region, 1.14 $\pm$ 0.87.
These results show that, when compared with the red GCs, the blue GCs have a larger velocity dispersion 
  and smaller rotation amplitude in the outer region.

\subsection{Comparison with Formation Models of Globular Clusters in gEs} \label{models}

The kinematic properties of the GC systems in galaxies can 
provide strong constraints on the formation models
  of the GC systems and their host galaxies \citep{cot01, cot03, hwa08, kor09}.
Several formation models of the GC systems in gEs have been suggested to describe the formation of GCs 
 \citep{pee69, ash92, har95, for97, cot98} and a summary of model descriptions and predictions can be found in several literature 
  \citep{rho01, lee03, ric04, wes04, bro06, hwa08}. 
These models can be broadly divided into four categories: 
  the monolithic collapse model, the major merger model, the multiphase
  dissipational collapse model, and the dissipationless accretion model.
These models suggest some kinematic properties of the GCs in galaxies.
These were discussed in comparison with the observation results for the GCs in gEs 
in \citet{hwa08} and will be summarized below. 
We focus on kinematic aspects of these models to compare with observational results below.
It is noted that these classical models  do not provide detailed kinematic properties
  so as to compare with the observational results.

In the monolithic collapse model an elliptical galaxy
and its GCs are formed through the collapse of an isolated massive
gas cloud or protogalaxy at high redshift \citep{lar75,car84,ari87}. 
In this model, 
the rotation of GCs can be generated by tidal torques from
companions \citep{pee69}, but the resulting rotation is not expected to be strong. 
This model cannot explain the strong rotation of the GC system seen in M60 and M87,
and the globally isotropic velocity ellipsoid of the GC system in several gEs.

In the major merger model  elliptical galaxies are formed by
a merger of two or more gas-rich disk galaxies \citep{too77,ash92,zep00}.
In this model, younger, spatially concentrated, red GCs are formed during the merger, while
spatially more extended, blue GCs come from the halos of the disk
galaxies (e.g., \citealt{bek02}). 
This model predicts that the newly formed red GCs show little
rotation compared to the blue GCs since the angular momentum would be
transported to the outer region during the merging process. 
This model cannot explain the presence of rotation of the red GCs in M60, M87, and NGC 5128,
while its prediction is consistent with the absence of rotation of the red GCs in other gEs.

In the multiphase dissipational collapse
model \citep{for97} elliptical galaxies form their GCs in distinct star formation
phases through a dissipational collapse, and they capture
some GCs by tidal effects from neighboring galaxies or the accretion of dwarf galaxies. 
The blue GCs are formed in the first star formation phase and the red GCs are
formed in the subsequent star formation phase after the gas in the
galaxy is self-enriched.
This model predicts that the blue GC system shows
no rotation and a high velocity dispersion, while the red GC
system shows some rotation depending on the degree of dissipation
in the collapse. 
Their prediction for the rotation is not consistent with the observational results for gEs in this study
(e.g., rotation measured in M60 and M87).

In the dissipationless accretion model \citep{cot98} 
the red GCs are formed in a dissipational monolithic collapse of a primary proto-galactic cloud, while the blue GCs
are subsequently captured from other (low-mass) galaxies through mergers or
tidal stripping. 
Since the blue GCs are captured from other  galaxies, 
they are expected to show a spatial distribution more extended than that of the red GCs. 
The blue GCs are expected to have radially biased orbits rather than isotropic or tangentially 
biased orbits, and are also expected to show no rotation \citep{ric04}. 
These predictions are not consistent with the observational results that the blue GCs in gEs show isotropic or tangential orbits as seen in Fig. \ref{fig-dispiso}(b),
 and that the blue GCs in four gEs show measurable rotation as seen in Table \ref{tab-gekin}.

Above classical models give some qualitative predictions for the kinematic properties of the GC systems,
but without any quantitative information.
On the other hand, numerical simulations provide  quantitative predictions 
for the kinematic properties of the GC systems in galaxies, which can be compared 
with the observational results. 
There are several numerical simulation studies that provided  
  some predictions of kinematic properties of the GC systems in galaxies 
  (e.g., \citealt{ves03,bek05,die05,kra05,moo06,bek08}).
Focusing on kinematic aspects of the GC systems in gEs,
  we compare our observational results for several gEs with the results in the simulations.

\citet{bek05} presented the results of a simulation of dissipationless major mergers of spiral galaxies
derived with an assumption that the spatial distribution of the GCs in E/S0's are initially similar to that
for the Milky Way Galaxy.
They predicted that both pre-existing metal-poor globular clusters (MPGCs) and metal-rich globular clusters (MRGCs)
  obtain stronger rotation in the outer region
  regardless of the orbital configuration of the merging galaxies.
In Figure \ref{fig-rotrad}, the blue GCs and red GCs in M87 and NGC 1407 show
slightly stronger rotation in the outer region  ($R>2R_{\rm eff}$)
  than in the inner region ($R<2R_{\rm eff}$), which is consistent with their prediction.
However, the blue GCs and red GCs in other gEs are not consistent with their prediction.

\citet{bek05}  also predicted that 
the velocity dispersion for both the MPGCs and MRGCs decreases 
as the galactocentric distance increases 
and that there is little difference between the MPGCs and MRGCs 
in all major merger models, while those are sometimes flat in  multiple merger models.
In addition the MPGCs should show slightly larger central velocity dispersion
than the MRGCs, indicating that the MPGCs are dynamically hotter than the MRGCs.
However, the observational results for NGC 4636 in Figure 5 and other gEs in Figure \ref{fig-disprad} show that there is significant difference in the VDPs between the MPGCs and MRGCs, which is not consistent with the prediction by \citet{bek05}.

Recently \citet{bek08} investigated the origin of GC systems using high-resolution cosmological
  N-body simulations combined with the semi-analytic models of galaxy formation.
They predicted that the majority (about 90 per cent) of GCs seen in the galaxy halos today 
  were formed in low-mass dwarf galaxies at redshifts larger than 3
  and that the mean formation epochs for the MPGCs ([Fe/H]$<-1$) and MRGCs ([Fe/H]$>-1$) 
  are, respectively, 12.7 Gyr ago (redshift $z=5.7$) and 12.3 Gyr ago (redshift $z=4.7$). 
MPGCs and MRGCs were also formed in the gas-rich major mergers for a wide range
  of redshifts, with peaks at $z\sim 5$ and $z\sim 3$, respectively.
MRGCs are formed also in isolated gas-rich galaxies.
They could explain several observational results on structural, kinematic and chemical properties 
  of the GC systems in various kinds of galaxies with their models. 

Kinematic properties of the GC systems predicted in the \citet{bek08}'s model are:
(a) the velocity dispersions of the MPGCs and MRGCs increase 
  according to the total luminosity of their host galaxy ($M_B$);
(b) the ratio of the velocity dispersions between the MPGCs and MRGCs
  is almost one for a wide range of the total luminosity of their host galaxy; and
(c) the ratio of maximum rotational velocities to central velocity dispersions in GC systems is low
  ($V/\sigma < 0.3$) for most galaxies, which is  in contrast to the case of disk-disk major mergers 
  leading to larger $V/\sigma$ ($> 0.5$)  suggested by \citet{bek05}.

Their prediction (a) is roughly consistent with the observational results shown 
  in Figure \ref{fig-kinmv}(c).
Their prediction (b) also appears to be consistent with our observational results
   considering the large uncertainties (see Fig. \ref{fig-kinmv}(f) in this study and Fig. 18 in \citealt{bek08}).
However, it is worth noting that
  the ratio of the velocity dispersions between the blue and red GCs ranges from 0.85 to 1.25 in Figure \ref{fig-kinmv}(f),
  that the ratio is the largest for the sample of the brightest and faintest galaxies,
  and that it is the smallest for the intermediately luminous galaxy.
This trend should be checked with a larger sample of galaxies.
Four out of seven gEs (NGC 4636, NGC 1399, M49, and NGC 5128)
  have low ratios of maximum rotational velocities to central velocity dispersions ($V/\sigma < 0.3$),
  as shown in Figure \ref{fig-kinmv}(g),
  which is again in a broad agreement with their prediction (c).

Therefore, the observational results appear to be approximately consistent
  with those in the simulations of \citet{bek08} in terms of the parameters that they considered.
However, it is noted that some GC systems show different behavior depending on the parameters
  and there remain many observational results to be explained (e.g., Figs. \ref{fig-kinmv} and \ref{fig-kinsn}). 

\subsection{A Mixture Scenario for the Origin of Globular Clusters in gEs} 

The most notable results emerging from our study of kinematic properties of the
GC systems in seven gEs are
(a) that the kinematic properties of the GC systems are diverse among gEs and
(b) that some kinematic parameters of the GC systems show strong correlations with the global
parameters of their host galaxies.
The first result indicates that the GCs in gEs
were probably formed and evolved via diverse mechanisms rather than via one single way,
and the second result implies that the kinematics of the GCs is controlled by the gravitational
potential of their host galaxies.
Considering all above comparisons and observational aspects of the GCs in gEs \citep{lee03, hwa08,pen08}, 
we derive a following scenario for the origin of GCs in gEs.

(1) MPGCs are formed mostly in low-mass dwarf galaxies very early, and preferentially in dwarf galaxies located in the high density environment like galaxy clusters. 
 These are the first generation of GCs in the universe. MPGCs should be also formed in massive galaxies as well, but the number of these massive galaxies is much smaller compared with that of the dwarf galaxies.
Observational evidence for this are: 
(a) MPGC are among the most metal-poor objects; 
(b) The spatial distribution of the MPGCs is more extended than that of the MRGCs around gEs \citep{lee98, dir05, lee03}; 
(c) The dwarfs in higher density environment have higher specific frequency than those in the lower density environments \citep{pen08};
and (d) the intracluster GCs are mostly MPGCs \citep{tam06}.  

(2) MRGCs are formed together with stars in massive galaxies or dissipational merging galaxies later than MPGCs, but not much later than MPGCs.  The chemical enrichment of the galaxies is rapid after the formation of MPGCs and the difference in the formation epoch of MPGCs and MRGCs should be small (e.g., much smaller than 1 Gyr).  
Observational evidence for this are: 
(a) MRGCs have, on average, similar metallicity to that of the stellar halo in gEs 
\citep{gei98,lee08b};
and (b) the difference in estimated ages between the MPGCs and MRGCs is small, 
while the difference in mean metallicity of the two populations of GCs is about a dex \citep{lee03,lee08b}.

(3) Massive galaxies grow becoming gEs via dissipationless or dissipational  merging of galaxies of various types and via accretion of many dwarf galaxies. New MRGCs will be formed during dissipational merging, but the faction of dissipational merging at this stage should be minor. 
A significant fraction of MPGCs in gEs we see today are from dissipationless merging or accretion.
Observational evidence for this are: 
(a) The spatial distribution of the MPGCs is more extended than that of the MRGCs around gEs \citep{lee98, lee03, dir05, lee08b}; 
(b) The intracluster GCs are mostly MPGCs \citep{tam06};
(c) The kinematics of the GCs are diverse among the gEs, and there are strong correlations
of the GC kinematics with some global parameters of their host galaxies found in this study. 
It is also  noted that \citet{kor09} concluded from the study of structure of  a large sample of elliptical galaxies that bright boxy elliptical galaxies (like gEs) were formed  via dissipationless (dry) merger, while faint disky
elliptical galaxies were formed via dissipational (wet) merger and conversion of late-type galaxies into spheroidals.

In this scenario each gE has a different history of growing involved with diverse merging and accretion,
explaining naturally the diversity in the kinematics of the GC systems in gEs. 
This scenario also explains the bimodal color distribution of the GCs, the difference in spatial distribution between the blue GCs and red GCs, the correlation  in color between the red GCs and their host  galaxies.
This is a mixture model or a {\it bibimbap} model ({\it bibimbap} is a Korean dish where diverse vegetables and warm rice are mixed together with some tasty sauces) in that it includes all the key elements in the previous models.

\section{Summary}\label{summary}

Using the photometric and spectroscopic database of 238 GCs (108 blue GCs and 130 red GCs) 
  in NGC 4636 (gE in the Virgo cluster), 
  we have derived the kinematics of the GC system of this galaxy.
Then we have compared the kinematics of the GC systems in seven gEs including NGC 4636, and
have investigated correlations between the GC kinematics and the globular parameters of their host galaxies. 
Our primary results are summarized as follows.

\begin{enumerate}

\item The red GC subsample of NGC 4636 shows marginal overall rotation.
In the inner region both the blue and red GC subsamples show some rotation.

\item Both of the velocity dispersions about the mean velocity and about the best fit rotation
curve of the blue GCs are about 50 km s$^{-1}$ larger than those of the red GCs. 

\item Comparison of observed stellar and GC velocity dispersion profiles with those
  calculated from the stellar mass profile shows that the mass-to-light ratio is not constant,
  but should  increase as the galactocentric distance increases,
  indicating the existence of an extended dark matter halo in NGC 4636.

\item  Using the X-ray mass profiles, 
  the number density distribution of GCs, and the observed VDP of GCs,  
  we have determined the velocity anisotropy of the NGC 4636 GC system.
The orbits of the NGC 4636 GC system are tangentially biased.
 The orbits of both the red GCs and blue GCs  at $R>20$ kpc are tangential, and the orbits of 
  the red GCs is slightly more tangential than that of the blue GCs.
The orbit of the blue GCs in the inner region at $R<12$ kpc is found to be radial.
  
\item From the  comparison of the kinematics of the NGC 4636 GC system in this study 
  with those for other gEs,
  we found several correlations between the kinematic properties of the GC systems 
  and the global parameters of their host galaxies.

\item We compared  the observational
  results for the GC systems in gEs including NGC 4636 
with those in several GC formation models, and found that 
some results are consistent with the predictions from the models
but some  are not. 
We suggested a mixture scenario for the origin of the GCs in gEs.

\end{enumerate}

\acknowledgments 
The authors are grateful to the anonymous referee for useful comments that improve the original
manuscript, and to the staff of the SUBARU Telescope
 for their kind help during the observation.
M.G.L. is supported in part by a grant (R01-2007-000-20336-0) 
from the Basic Research Program of the
Korea Science and Engineering Foundation. 
N.A is financially supported in part by a Grant-in-Aid for
Scientific Research by the Japanese Ministry of Education, Culture,
Sports, Science and Technology (No. 19540245).


\clearpage

\begin{deluxetable}{cccrrrrrr}
\tablewidth{0pc} 
\tablecaption{Kinematics of the NGC 4636 Globular Cluster System\label{tab-n4636kin}}
\tablehead{
\colhead{$R$} & \colhead{$\langle R\rangle$} & \colhead{} &
\colhead{$\overline{v_p}$} & \colhead{$\sigma_p$} &
\colhead{$\Theta_0$} & \colhead{$\Omega R$} &
\colhead{${\sigma}_{p,r}$} & \colhead{}  \\
\colhead{(arcsec)} & \colhead{(arcsec)} & \colhead{$N$} & 
\colhead{(km s$^{-1}$)} & \colhead{(km s$^{-1}$)} & \colhead{(deg)} & \colhead{(km s$^{-1}$)} & 
\colhead{(km s$^{-1}$)} & \colhead{$\Omega R$/${\sigma}_{p,r}$} \\  
\colhead{(1)} & \colhead{(2)} & \colhead{(3)} & \colhead{(4)} & 
\colhead{(5)} & \colhead{(6)} & \colhead{(7)} & \colhead{(8)} & \colhead{(9)}}

\startdata
\multicolumn{9}{c}{All GCs: 238 Clusters with 0.9 $\le$ $(C-T_1)$ $<$ 2.1} \\ 
\hline
  23$-$ 926& 268& 238 & $  949_{- 16}^{+ 13}$ & $  225_{-  9}^{+ 12}$ & $  174_{- 48}^{+ 73}$ & $   37_{- 30}^{+ 32}$ & $  226_{-  9}^{+ 12}$ & $0.16_{-0.14}^{+0.14}$ \\
  23$-$ 259& 150& 120 & $  951_{- 27}^{+ 27}$ & $  217_{- 11}^{+ 13}$ & $  163_{- 64}^{+ 87}$ & $   42_{- 34}^{+ 32}$ & $  220_{- 10}^{+ 15}$ & $0.19_{-0.16}^{+0.15}$\\
 260$-$ 926& 388& 118 & $  946_{- 17}^{+ 20}$ & $  234_{- 17}^{+ 16}$ & $  195_{- 64}^{+ 59}$ & $   31_{- 21}^{+ 27}$ & $  234_{- 17}^{+ 18}$ & $0.13_{-0.09}^{+0.13}$\\
\hline
\multicolumn{9}{c}{Blue GCs: 108 Clusters with 0.9 $\le$ $(C-T_1)$ $<$ 1.55} \\
\hline  
  23$-$ 926& 298& 108 & $  951_{- 25}^{+ 30}$ & $  251_{- 12}^{+ 18}$ & $    0_{-144}^{+146}$ & $   27_{- 24}^{+ 34}$ & $  252_{- 15}^{+ 15}$ & $0.11_{-0.09}^{+0.13}$\\
  23$-$ 252& 145&  42 & $  943_{- 44}^{+ 33}$ & $  244_{- 29}^{+ 26}$ & $    5_{- 22}^{+299}$ & $  119_{- 62}^{+ 72}$ & $  253_{- 39}^{+ 37}$ & $0.47_{-0.24}^{+0.30}$\\
 267$-$ 926& 395&  66 & $  954_{- 29}^{+ 42}$ & $  257_{- 28}^{+ 30}$ & $  235_{-103}^{+ 84}$ & $   18_{- 33}^{+ 34}$ & $  259_{- 26}^{+ 32}$ & $0.07_{-0.12}^{+0.14}$\\
\hline
 \multicolumn{9}{c}{Red GCs: 130 Clusters with 1.55 $\le$ $(C-T_1)$ $<$ 2.1} \\
\hline
  25$-$ 713& 243& 130 & $  949_{- 19}^{+ 20}$ & $  205_{- 13}^{+ 11}$ & $  178_{- 34}^{+ 53}$ & $   68_{- 35}^{+ 48}$ & $  203_{- 13}^{+ 12}$ & $0.33_{-0.17}^{+0.21}$\\
  25$-$ 259& 152&  78 & $  954_{- 25}^{+ 29}$ & $  204_{- 12}^{+ 13}$ & $  179_{- 49}^{+ 56}$ & $   77_{- 47}^{+ 44}$ & $  207_{- 12}^{+ 14}$ & $0.37_{-0.19}^{+0.23}$\\
 260$-$ 713& 380&  52 & $  940_{- 26}^{+ 37}$ & $  206_{- 23}^{+ 28}$ & $  104_{- 57}^{+ 90}$ & $    0_{- 34}^{+ 41}$ & $  206_{- 22}^{+ 32}$ & $0.00_{-0.17}^{+0.21}$
\enddata
\end{deluxetable}

\begin{deluxetable}{cccrrrrrr}
\tablewidth{0pc} 
\tablecaption{Kinematics of the NGC 1407 Globular Cluster System\label{tab-n1407kin}}
\tablehead{
\colhead{$R$} & \colhead{$\langle R\rangle$} & \colhead{} &
\colhead{$\overline{v_p}$} & \colhead{$\sigma_p$} &
\colhead{$\Theta_0$} & \colhead{$\Omega R$} &
\colhead{${\sigma}_{p,r}$} & \colhead{}  \\
\colhead{(arcsec)} & \colhead{(arcsec)} & \colhead{$N$} & 
\colhead{(km s$^{-1}$)} & \colhead{(km s$^{-1}$)} & \colhead{(deg)} & \colhead{(km s$^{-1}$)} & 
\colhead{(km s$^{-1}$)} & \colhead{$\Omega R$/${\sigma}_{p,r}$} \\  
\colhead{(1)} & \colhead{(2)} & \colhead{(3)} & \colhead{(4)} & 
\colhead{(5)} & \colhead{(6)} & \colhead{(7)} & \colhead{(8)} & \colhead{(9)}}

\startdata
\multicolumn{9}{c}{All GCs: 172 Clusters with 0.70 $<$ $(g'-i')_0$ $<$ 1.30} \\ 
\hline
21 $-$ 666 & 247 & 172 & $1753^{+19}_{-18}~$ & $245^{+12}_{-12}~$ & $226^{+21}_{-28}~$ & $86^{+27}_{-35}~$ & $247^{+12}_{-13}~$ & $0.35^{+0.11}_{-0.12}$ \\ 
21 $-$ 221 & 126 & 84 & $1723^{+30}_{-28}~$ & $249^{+22}_{-22}~$ & $225^{+36}_{-50}~$ & $63^{+41}_{-35}~$ & $256^{+27}_{-22}~$ & $0.25^{+0.17}_{-0.15}$ \\ 
226 $-$ 666 & 364 & 88 & $1781^{+31}_{-25}~$ & $238^{+17}_{-16}~$ & $230^{+17}_{-23}~$ & $114^{+36}_{-37}~$ & $232^{+18}_{-18}~$ & $0.49^{+0.16}_{-0.17}$ \\ 
\hline
\multicolumn{9}{c}{Blue GCs: 86 Clusters with 0.70 $<$ $(g'-i')_0$ $\le$ 0.98} \\
\hline 
 28 $-$ 663 & 272 & 86 & $1757^{+34}_{-35}~$ & $240^{+19}_{-24}~$ & $271^{+29}_{-27}~$ & $87^{+35}_{-29}~$ & $237^{+21}_{-22}~$ & $0.36^{+0.15}_{-0.13}$ \\ 
28 $-$ 221 & 136 & 36 & $1726^{+36}_{-33}~$ & $213^{+19}_{-18}~$ & $254^{+81}_{-95}~$ & $42^{+36}_{-41}~$ & $221^{+23}_{-26}~$ & $0.19^{+0.18}_{-0.19}$ \\ 
226 $-$ 663 & 369 & 50 & $1781^{+47}_{-39}~$ & $259^{+26}_{-23}~$ & $268^{+27}_{-16}~$ & $151^{+49}_{-41}~$ & $238^{+33}_{-25}~$ & $0.64^{+0.26}_{-0.20}$ \\ 
\hline
 \multicolumn{9}{c}{Red GCs: 86 Clusters with 0.98 $<$ $(g'-i')_0$ $<$ 1.30} \\
\hline
21 $-$ 666 & 223 & 86 & $1750^{+29}_{-22}~$ & $251^{+19}_{-21}~$ & $209^{+27}_{-31}~$ & $104^{+38}_{-36}~$ & $250^{+17}_{-21}~$ & $0.41^{+0.17}_{-0.15}$ \\ 
21 $-$ 217 & 118 & 48 & $1723^{+43}_{-44}~$ & $278^{+36}_{-30}~$ & $223^{+40}_{-50}~$ & $80^{+55}_{-54}~$ & $282^{+39}_{-32}~$ & $0.28^{+0.23}_{-0.18}$ \\ 
227 $-$ 666 & 357 & 38 & $1781^{+40}_{-39}~$ & $214^{+25}_{-20}~$ & $189^{+23}_{-23}~$ & $150^{+50}_{-47}~$ & $199^{+27}_{-19}~$ & $0.75^{+0.29}_{-0.25}$ \\ 
\enddata
\end{deluxetable}
\clearpage
\begin{deluxetable}{ccrrcrrcccc}
\tabletypesize{\scriptsize}
\tablewidth{0pc} 
\tablecaption{Giant Elliptical Galaxy Samples\label{tab-gesample}}
\tablehead{ 
\colhead{Galaxy} & \colhead{$M_V$\tablenotemark{~a}} &
\colhead{$\upsilon_{\rm sys}$\tablenotemark{~b}} &
\colhead{R$_{\rm eff}$\tablenotemark{~c}} &
\colhead{$\epsilon$\tablenotemark{~d}} &
\colhead{P.A.$_{\rm min}$\tablenotemark{~e}} &
\colhead{Distance\tablenotemark{~f}} &
\colhead{$\sigma_{star}$\tablenotemark{~g}} &
\colhead{log($L_X$)\tablenotemark{~h}} &
\colhead{N$_{GC}$\tablenotemark{~i}} &
\colhead{S$_{N}$\tablenotemark{~j}} \\
\colhead{} & \colhead{} & \colhead{(km s$^{-1}$)} &
\colhead{(kpc)} & & \colhead{(deg)} & \colhead{(Mpc)} &
\colhead{(km s$^{-1}$)} & \colhead{(erg s$^{-1}$)} &
\colhead{blue red} & \colhead{} \\
\colhead{(1)} & \colhead{(2)} & \colhead{(3)} & \colhead{(4)} & \colhead{(5)} & 
\colhead{(6)} & \colhead{(7)} & \colhead{(8)} & \colhead{(9)} & \colhead{(10)} & \colhead{(11)}}

\startdata
M60      & $-$22.13 & 1056 & 7.33 & 0.216 &  15 & 16.8 & 337$\pm$~9 & 41.27$\pm$0.042 & 663 455 &  4.0$\pm$0.8 \\
M87      & $-$22.38 & 1307 & 7.16 & 0.125 &  69 & 16.1 & 330$\pm$~5 & 43.08$\pm$0.007 & 928 771 & 14.1$\pm$1.5 \\
M49      & $-$22.57 & ~997 & 9.48 & 0.175 &  65 & 16.3 & 294$\pm$~5 & 41.71$\pm$0.043 & 683 563 &  3.6$\pm$0.6 \\
NGC 1399 & $-$21.71 & 1442 & 14.55 & 0.099 & 20 & 20.0 & 250$\pm$12 & 42.18$\pm$0.040 & 459 500 &  5.1$\pm$1.2 \\
NGC 5128 & $-$21.90 & ~541 & 6.02 & 0.224 & 125 &  4.2 & 140$\pm$40 & 40.15$\pm$0.200 & 178 158 &  1.8$\pm$0.5 \\
NGC 4636 & $-$21.67 & ~928 & 6.35 & 0.256 &  58 & 14.7 & 202$\pm$~3 & 41.68$\pm$0.046 & 639 633 &  8.2$\pm$1.6 \\
NGC 1407 & $-$21.86 & 1784 & 7.34 & 0.050 & 148 & 21.0 & 265$\pm$15 & 41.14$\pm$0.092 & 446 725 &  3.8$\pm$1.3\\
\enddata

\tablenotetext{a~}{$V$-band absolute total magnitude: NGC 5128 \citep{duf79}, NGC 1407 \citep{ben92}, and other galaxies \citep{fab97}.}
\tablenotetext{b~}{Systemic velocity: M60 \citep{lee08b}, M87, M49 \citep{smi00},
  NGC 1399 \citep{ric04}, NGC 5128 \citep{hui95}, NGC 4636 (Paper I), and NGC 1407 \citep{rom09}.}
\tablenotetext{c~}{Effective radius in units of kpc: M60 \citep{lee08b}, NGC 5128 \citep{duf79}, NGC 1407 \citep{spo08}, and other galaxies \citep{kim06}.}
\tablenotetext{d~}{Ellipticity: NGC 5128 \citep{dev91}, NGC 1407 \citep{spo08}, and other galaxies \citep{kim06}.}
\tablenotetext{e~}{Position angle of the minor axis: M60 \citep{lee08b}, NGC 1399 \citep{sag00}, 
  NGC 5128 \citep{duf79}, NGC 1407 \citep{spo08}, and other galaxies \citep{kim06}.}
\tablenotetext{f~}{Distance in units of Mpc: NGC 1407 \citep{spo08}, and other galaxies \citet{ton01}.}
\tablenotetext{g~}{Mean velocity stellar dispersion at $\sim R_{\rm eff}/4$: NGC 1399 \citep{sag00}, NGC 5128 \citep{wil86}, NGC 1407 \citep{spo08}, and other galaxies \citep{ben94} .}
\tablenotetext{h~}{ { Logarithmic} value of X-ray luminosity: NGC 5128 \citep{osu01, osu03}, and other galaxies \citep{beu99}. }
\tablenotetext{i~}{Numbers of blue GCs and red GCs: NGC 5128 \citep{woo07}, NGC 1407 \citep{for06}, and other galaxies \citep{kim06}.}
\tablenotetext{j~}{Specific frequency of GCs: M60 \citep{lee08b, for04}, NGC 5128 \citep{har04, har06}, NGC 4636 \citep{kis94, dir05}, NGC 1407 \citep{per97, for06}, and other galaxies \citep{bro06}.}
\end{deluxetable}
\clearpage
\begin{deluxetable}{ccrrrrrcc}
\tablewidth{0pc} 
\tablecaption{Global Kinematic Properties of GCs in gEs\label{tab-gekin}}
\tablehead{ \colhead{Galaxy} & \colhead{GC} & \colhead{$N$} &
\colhead{$\overline{v_p}$} & \colhead{$\sigma_p$} &
\colhead{$\Theta_0$} & \colhead{$\Omega R$} &
\colhead{${\sigma}_{p,r}$} & \colhead{$\Omega R$/${\sigma}_{p,r}$} \\
\colhead{} & \colhead{} & \colhead{} & \colhead{(km s$^{-1}$)} & \colhead{(km s$^{-1}$)} &
\colhead{(deg)} & \colhead{(km s$^{-1}$)} & \colhead{(km s$^{-1}$)} & \colhead{} }
\startdata
M60 & AGC & 121& $1073^{+22}_{-22}$ & $234^{+13}_{-14}$ & $225^{+12}_{-14}$ & $141^{+50}_{-38}~$ & $217^{+14}_{-16}$ & $0.65^{+0.27}_{-0.22}$  \\
    & BGC &  83& $1086^{+27}_{-25}$ & $223^{+13}_{-16}$ & $218^{+16}_{-23}$ & $130^{+62}_{-51}~$ & $207^{+15}_{-19}$ & $0.63^{+0.35}_{-0.30}$  \\
    & RGC &  38& $1040^{+48}_{-42}$ & $258^{+21}_{-31}$ & $237^{+18}_{-19}$ & $171^{+58}_{-46}~$ & $240^{+20}_{-34}~$ & $0.71^{+0.30}_{-0.29}$  \\
\multicolumn{9}{c}{} \\
M87 & AGC & 276& $1333^{+ 25}_{-23}$&$  414^{+ 15}_{-18}$&$   68^{+ 11}_{-12}$&$  172^{+ 39}_{-28}$&$  399^{+ 15}_{-18}$&$  0.43^{+0.12}_{-0.09}$ \\
    & BGC & 158& $1341^{+ 36}_{-33}$&$  425^{+ 22}_{-25}$&$   59^{+ 17}_{-17}$&$  181^{+ 57}_{-44}$&$  414^{+ 22}_{-26}$&$  0.44^{+0.16}_{-0.13}$ \\
    & RGC & 118& $1324^{+ 39}_{-37}$&$  400^{+ 25}_{-28}$&$   79^{+ 16}_{-17}$&$  165^{+ 53}_{-33}$&$  380^{+ 24}_{-27}$&$  0.43^{+0.17}_{-0.12}$ \\
\multicolumn{9}{c}{} \\
M49 & AGC & 263& $ 973^{+ 20}_{-18}$&$  322^{+ 14}_{-17}$&$  106^{+ 44}_{-45}$&$   54^{+ 50}_{-23}$&$  321^{+ 14}_{-17}$&$  0.17^{+0.17}_{-0.08}$ \\
    & BGC & 159& $ 954^{+ 32}_{-27}$&$  352^{+ 21}_{-25}$&$  102^{+ 36}_{-37}$&$   92^{+ 71}_{-35}$&$  349^{+ 21}_{-24}$&$  0.27^{+0.22}_{-0.12}$ \\
    & RGC & 104& $ 999^{+ 31}_{-25}$&$  276^{+ 19}_{-23}$&$  182^{+ 53}_{-50}$&$   11^{+ 79}_{-83}$&$  278^{+ 19}_{-23}$&$  0.04^{+0.29}_{-0.30}$ \\
\multicolumn{9}{c}{} \\
NGC 1399 & AGC & 435& $1442^{+ 15}_{-14}$&$  323^{+ 11}_{-13}$&$  307^{+ 50}_{-46}$&$   31^{+ 43}_{-48}$&$  326^{+ 11}_{-13}$&$  0.10^{+0.14}_{-0.15}$ \\
         & BGC & 216& $1445^{+ 26}_{-22}$&$  359^{+ 17}_{-21}$&$  261^{+ 45}_{-52}$&$   69^{+ 68}_{-29}$&$  364^{+ 18}_{-21}$&$  0.19^{+0.20}_{-0.09}$ \\
         & RGC & 219& $1439^{+ 19}_{-17}$&$  285^{+ 16}_{-19}$&$    0^{+ 40}_{-36}$&$   46^{+ 53}_{-39}$&$  288^{+ 16}_{-19}$&$  0.16^{+0.19}_{-0.15}$ \\
\multicolumn{9}{c}{} \\
NGC 5128 & AGC & 341& $ 536^{+  9}_{ -8}~$&$  129^{+  5}_{ -7}~$&$  184^{+ 23}_{-26}$&$   30^{+ 16}_{-14}$&$  129^{+  5}_{ -7}~$&$  0.23^{+0.14}_{-0.13}$ \\
         & BGC & 160& $ 526^{+ 11}_{-11}$&$  126^{+  7}_{ -8}~$&$  168^{+ 27}_{-47}$&$   25^{+ 22}_{-35}$&$  129^{+  7}_{ -7}~$&$  0.19^{+0.18}_{-0.29}$ \\
         & RGC & 146& $ 552^{+ 16}_{-15}$&$  133^{+  9}_{-11}$&$  191^{+ 40}_{-75}$&$   47^{+ 43}_{-54}$&$  132^{+  9}_{-11}$&$  0.36^{+0.35}_{-0.44}$ \\
\multicolumn{9}{c}{} \\
NGC 4636 & AGC &  238& $ 949^{+ 13}_{-16}$&$  225^{+ 12}_{-9}$&$  174^{+ 73}_{-48}$&$   37^{+ 32}_{-30}$&$  226^{+ 12}_{-9}$&$  0.16^{+0.14}_{-0.14}$ \\
         & BGC &  108& $ 951^{+ 30}_{-25}$&$  251^{+ 18}_{-12}$&$  0^{+ 146}_{-144}$&$   27^{+ 34}_{-24}$&$  252^{+ 15}_{-17}$&$  0.11^{+0.13}_{-0.09}$ \\
         & RGC &  130& $ 949^{+ 20}_{-19}$&$  205^{+ 11}_{-13}$&$  178^{+53}_{-34}$&$   68^{+ 48}_{-35}$&$  203^{+ 12}_{-13}$&$  0.33^{+0.21}_{-0.17}$ \\
\multicolumn{9}{c}{} \\      
NGC 1407 & AGC & 172 & $1753^{+19}_{-18}~$ & $245^{+12}_{-12}~$ & $226^{+21}_{-28}~$ & $86^{+27}_{-35}~$ & $247^{+12}_{-13}~$ & $0.35^{+0.11}_{-0.12}$ \\
         & BGC & 86 & $1757^{+34}_{-35}~$ & $240^{+19}_{-24}~$ & $271^{+29}_{-27}~$ & $87^{+35}_{-29}~$ & $237^{+21}_{-22}~$ & $0.36^{+0.15}_{-0.13}$ \\
         & RGC & 86 & $1750^{+29}_{-22}~$ & $251^{+19}_{-21}~$ & $209^{+27}_{-31}~$ & $104^{+38}_{-36}~$ & $250^{+17}_{-21}~$ & $0.41^{+0.17}_{-0.15}$ \\
\enddata
\end{deluxetable}
\clearpage

\begin{deluxetable}{cccrrccl}
\tablewidth{0pc}
\tablecaption{Correlations between the GC systems and their Host Galaxies$^a$ \label{tab-relation}}
\tablehead{
\colhead{X} &
\colhead{Y} &
\colhead{$a$} &
\colhead{$b$} &
\colhead{rms} & \colhead{$r_S$} & \colhead{$\sigma(r_S )$} & 
\colhead{Remarks} }
\startdata
 $\log L_X$       & $\sigma_{p,r}^{AGC}$ & 97.66 & --3796 & 27.77 & 0.89 & 0.01& Fig. \ref{fig-kinmv}(a) \\ 
 $ \log L_X$       & $\sigma_{p,r}^{BGC}$ & 110.28 & --4309 & 32.48 & 0.96 & 0.00& Fig. \ref{fig-kinmv}(a) \\ 
 $ \log L_X$       & $\sigma_{p,r}^{RGC}$ & 84.04 & --3243 & 27.13 & 0.86 & 0.01& Fig. \ref{fig-kinmv}(a) \\ 
 $\sigma_{star}$ & $\sigma_{p,r}^{AGC}$ & 1.394 & --69.51 & 30.71 &0.83 & 0.04 & Fig. \ref{fig-kinmv}(b) \\ 
 $\sigma_{star}$ & $\sigma_{p,r}^{BGC}$ & 1.538 & --88.77 & 45.63 &0.71 & 0.11 & Fig. \ref{fig-kinmv}(b) \\ 
 $\sigma_{star}$ & $\sigma_{p,r}^{RGC}$ & 1.237 & --50.31 & 23.93 &0.83 & 0.04 & Fig. \ref{fig-kinmv}(b) \\ 
 $ M_V$          & $\sigma_{p,r}^{AGC}$ & --255.6 & -5389 & 64.80 & --0.49 & 0.33 & Fig. \ref{fig-kinmv}(c) \\ 
 $ M_V$          & $\sigma_{p,r}^{BGC}$ & --279.0 & -5898 & 71.90 & --0.43 & 0.40 & Fig. \ref{fig-kinmv}(c) \\ 
 $ M_V$          & $\sigma_{p,r}^{RGC}$ & --222.1 & -4658 & 59.60 & --0.66 & 0.16 & Fig. \ref{fig-kinmv}(c) \\ 
 $S_N$           & $\sigma_{p,r}^{AGC}$ & 20.03 & 150.1 & 63.28 & 0.57 & 0.18 & Fig. \ref{fig-kinsn}(a) \\ 
$\epsilon$  & $\sigma_{p,r}^{AGC}$ & -1548 & 552.2 & 56.70 & --0.77 & 0.07 & Fig. \ref{fig-kinsn}(c) \\ 
$N_{BGC} / N_{RGC}$ & $\sigma_{p,r}^{BGC}/\sigma_{p,r}^{RGC}$& --0.908 & 2.163 & 0.107 & --0.60 & 0.21 & Fig. \ref{fig-kinsn}(e) \\ 
$N_{BGC} / N_{RGC}$ & $\Omega R /\sigma_{p,r}^{AGC}$  & 1.120 & --1.003 & 0.091 & 0.83 & 0.04 & Fig. \ref{fig-kinsn}(h) \\ 
$\sigma_{p,r}^{BGC}/\sigma_{p,r}^{RGC}$ & $\Omega R /\sigma_{p,r}^{AGC}$ & --1.153 & 1.556 & 0.108 & --0.86 & 0.01 & Fig. \ref{fig-disprot}(a) 
\enddata
\tablenotetext{a} { {L}inear fits with $Y=aX+b$ using the bisector method \citep{iso90}. $r_S$ and $\sigma(r_S )$ represent the
Spearman's rank coefficient and its significance, respectively.}
\end{deluxetable}
\clearpage

\clearpage

\clearpage

\clearpage

\end{document}